\documentclass[]{article}
\usepackage{txfonts}
\usepackage{graphicx}
\usepackage{natbib}
\bibpunct{(}{)}{;}{a}{}{,} % to follow the A&A style

\newcommand{\hs}{\hspace{0.08cm}}

\begin{document}

%--------------------------------------------------------------------------
% Title
%--------------------------------------------------------------------------
\title{\bf Spectropolarimetric signatures \\
       of Earth--like extrasolar planets}

\author{D. M. Stam (d.m.stam@sron.nl) \\ \\
       Aerospace Engineering Department, Technical University Delft, \\ 
           Kluyverweg 1, 2629 HS, Delft, The Netherlands \\ and \\
           SRON Netherlands Institute for Space Research, \\
           Sorbonnelaan 2, 3584 CA Utrecht, The Netherlands}
	   
\maketitle

%\date{Received xxx / Accepted yyy}
                                                                                
%--------------------------------------------------------------------------
% Abstract
%--------------------------------------------------------------------------
\abstract{
We present results of numerical simulations of the flux (irradiance),
$F$, and the degree of polarization (i.e. the ratio of polarized
to total flux), $P$, of light that is reflected by Earth--like 
extrasolar planets 
orbiting solar--type stars, as functions of the wavelength 
(from 0.3 to 1.0~$\mu$m, with 0.001~$\mu$m spectral resolution)
and as functions of the planetary phase angle.
We use different surface coverages for our model planets,
including vegetation and a Fresnel reflecting ocean, and
clear and cloudy atmospheres.
Our adding-doubling radiative transfer algorithm, which fully includes 
multiple scattering and polarization, handles horizontally homogeneous
planets only; we simulate fluxes and polarization of 
horizontally inhomogeneous planets by weighting results for
homogeneous planets.
Like the flux, $F$, the degree of polarization, $P$, of the reflected 
starlight is shown to depend strongly on the phase angle, 
on the composition and structure of the planetary atmosphere, on
the reflective properties of the underlying surface, and on 
the wavelength, in particular in wavelength regions with gaseous
absorption bands.
The sensitivity of $P$ to a planet's physical properties appears to
be different than that of $F$. Combining flux with polarization
observations thus makes for
a strong tool for characterizing extrasolar planets.
The calculated total and polarized fluxes will be made available 
through the CDS. \\

keywords: techniques: polarimetric --
          stars: planetary systems --
          polarization}
                                                                                
%\titlerunning{Spectropolarimetry of Earth-like exoplanets}
%\authorrunning{Stam}

%-----------------------------------------------------------------------
% Section 1
%-----------------------------------------------------------------------
\section{Introduction}
\label{sect_introduction}

Polarimetry has been recognized as a powerful technique for 
enhancing the contrast between a star and an exoplanet, and 
hence for the direct detection of exoplanets,
because, integrated over the stellar disk, the direct light of
a solar type star can be considered to be unpolarized 
\citep[see][]{1987Natur.326..270K},
while the starlight that has been reflected by a planet will generally
be polarized because it has been scattered within the planetary
atmosphere and/or reflected by the surface (if there is any).
The degree of polarization of the reflected starlight 
(i.e. the ratio of the polarized to the total flux)
is expected to be especially large around a planet's quadrature 
(i.e. when the planet is seen at a phase angle of 90$^\circ$), 
where the angular separation between a star and its exoplanet
is largest, and which is thus an excellent phase angle for 
the direct detection of light from an exoplanet.

Besides from detecting exoplanets,
polarimetry can also be used to characterize exoplanets,
because the planet's degree of polarization as a function of wavelength 
and/or planetary phase angle is sensitive to the structure and 
composition of the planetary atmosphere and underlying surface.
This application of polarimetry is well-known from remote-sensing 
of solar system planets, in particular Venus 
\citep[see][for a classic example]{1974JAtS...31.1137H,
1974IAUS...65..197H}, 
but also the outer planets (see \citet{2005Icar..176....1S}
for recent, Hubble Space Telescope polarization observations of Mars,
and \citet{2005ASPC..343..189J} and \citet{2006A&A...452..657S}  
for Earth-based polarimetry of Uranus and Neptune).
Note that Venus is much more favourable to 
observe with Earth-based polarimetry than the outer solar system 
planets, because as an inner planet, Venus can be observed from 
small to large phase 
angles (including quadrature), whereas the outer planets are always
seen at small phase angles, where the observed light is mostly
backscattered light and degrees of polarization are thus usually
small (see \citet{2004A&A...428..663S} for examples of the 
phase angle dependence of the degree of polarization of starlight
reflected by gaseous exoplanets).

The strengths of polarimetry for exoplanet detection and
characterization have been recognized and described before, 
for example by \citet{2000ApJ...540..504S,2003ASPC..294..529S,
2003toed.conf...11H,2003toed.conf..615S,2003ASPC..294..535S,
2004A&A...428..663S,2005ASPC..343..207S},
who presented numerically calculated fluxes and
degrees of polarization of gaseous exoplanets. 
Note that \citet{2000ApJ...540..504S,2003ASPC..294..529S}, and
\citet{2003toed.conf...11H}
concentrate on polarization
signals of exoplanets that are spatially {\em un}resolvable 
from their star, in other words, the polarized flux of the 
planet is added to a huge background of unpolarized stellar
flux, while \citet{2003toed.conf..615S} and 
\citet{2003ASPC..294..535S,2004A&A...428..663S,2005ASPC..343..207S} 
aim at spatially
resolvable planets, which are observed with a significantly
smaller unpolarized, stellar background signal.
Polarization signals of spatially unresolved {\em non-spherical}
planets were presented by \citet{2006ApJ...639.1147S}.
Note that their calculations include only single scattered
light, and not all orders of scattering (like those
of \citet{2000ApJ...540..504S} and \citet{2004A&A...428..663S}), which,
except for planetary atmospheres with a very thin scattering
or a very thick absorption optical thickness,
significantly influences the predicted degree of polarization,
because multiple scattered light usually has a (much) 
lower degree of polarization than singly scattered light.

Examples of ground-based telescope instruments that use 
polarimetry for exoplanet research are PlanetPol, which
aims at detecting spatially unresolved gaseous exoplanets
\citep[see][and references therein]{2006SPIE.6269E..25H,
2006PASP..118.1305H} and SPHERE (Spectro-Polarimetric 
High-contrast Exoplanet Research) 
\citep[see][and references therein]{2006Msngr.125...29B} 
which aims at detecting spatially resolved gaseous exoplanets. 
SPHERE is being designed and build for ESO's Very Large 
Telescope (first light is expected in 2010) and has a 
polarimeter based on the ZIMPOL
(Z\"{u}rich Imaging Polarimeter) technique
\citep[see][and references therein]{2005ASPC..343...89S,2004SPIE.5492..463G}.
Polarimetry is also a technique used in SEE-COAST (the Super Earths
Explorer -- Coronographic Off-Axis Telescope), a space-based
telescope for the detection and the characterization
of gaseous exoplanets and large rocky exoplanets, so called
'Super-Earths',
\citep[][]{2006sf2a.conf..429S}, that has been proposed
to ESA in response to its 2007 Cosmic Vision call.

The design and development of instruments for the direct 
detection of (polarized) light of exoplanets requires 
sample signals, i.e. total and polarized fluxes as
functions of the wavelength and as functions of the 
planetary phase angle.
Previously \citep[see e.g.][]{2004A&A...428..663S}, we presented 
numerically calculated flux and polarization spectra of light 
reflected by {\em giant, gaseous} exoplanets, integrated over
the illuminated and visible part of the planetary disk, 
for various phase angles.
In this paper, we present similar spectra
but now for light reflected by {\em Earth--like} exoplanets.
Our radiative transfer calculations fully include single and
multiple scattering and polarization.
The model atmospheres 
contain either only gaseous molecules or a combination of gas
and clouds. The clouds are modeled as horizontally homogeneous
layers of scattering (liquid water) particles, which allows 
surface features to show up in the reflected light even 
if the planet is fully covered. 
We show results for surfaces with wavelength
independent albedos ranging from 0.0 to 1.0, as well as
for surface albedos representative for vegetation, and
ocean. The ocean surface includes Fresnel reflection.

Our disk integration method is based on the expansion of
the radiation field of the planet into generalized spherical
functions \citep[][]{2006A&A...452..669S}, and pertains to
horizontally homogeneous planets only (the planetary atmospheres
can be vertically inhomogeneous). The main advantage of this
method, compared to more conventional integration of calculated
fluxes and polarization over a planetary disk
is that the flux and polarization of a planet can be
rapidly obtained for an arbitrary number of planetary
phase angles, without the need of new radiative transfer 
calculations for every new phase angle. 
This is indeed an important advantage, because polarization 
calculations are generally very computing time consuming 
compared to mere flux calculations.
The disadvantage of our method is obviously its inability to
handle horizontally inhomogeneous planets.
In this paper, we will approximate the light reflected by 
horizontally inhomogeneous planets by using weighted sums 
of light reflected by
horizontally homogeneous planets. With such {\em quasi} horizontally
inhomogeneous planets, we can still get a good impression of 
the influence of horizontal inhomogeneities on the reflected
signals. When in the future direct observations
of Earth-like exoplanets become available, the more conventional
disk integration method can straightforwardly be applied.

Our numerical simulations cover the wavelength region 
from 0.3 to 1.0~$\mu$m, thus from the UV to the near-infrared.
The spectral resolution of our simulations is 0.001~$\mu$m,
which is high enough for spectral features due to absorption 
of atmospheric gases to be clearly visible in the flux and 
polarization spectra.
Such high spectral resolution observations of Earth-like
exoplanets will not be possible for years to come; 
our spectra, however, show the potential information 
content of high spectral resolution spectra, and they can
be convolved with instrument response functions 
to simulate observations by instruments with a lower spectral
resolution. To allow the use of our flux and polarization 
spectra for such applications,
they will be made available at the CDS.

Flux spectra of light reflected by Earth-like exoplanets
have been presented before \citep[e.g. by][]{2006AsBio...6..881T,
2006AsBio...6...34T,
2006ApJ...644L.129T,2006ApJ...651..544M,2006ApJ...644..551T}.
New in this paper are flux spectra with the corresponding 
polarization spectra. 
Numerically calculated degrees and directions of polarization 
of exoplanets are not only useful for the design and building 
of polarimeters for exoplanet research, as described above,
but also for the design, building, and use
of instruments that aim at measuring only fluxes of exoplanets.
Namely, unless carefully corrected for, the optical components
of such instruments will
be sensitive to the state of polarization of the observed light.
Consequently, the measured fluxes will depend on the state of
polarization of the observed light.
Provided an instrument's polarization sensitivity is known,
our simulations can help to {\em estimate} the error that arises
in the measured fluxes. Note that in order to actually
{\em correct} measured fluxes for instrumental polarization sensitivity, 
knowing the 
polarization sensitivity of an instrument does not suffice;
the state of polarization of the incoming light should be 
measured along with the flux
\citep[see e.g.][for a discussion on flux errors in
remote-sensing due to
instrumental polarization sensitivity]{2000JGR...10522379S}.

\citet{2005A&A...444..275S} discuss another reason to 
include polarization into numerical simulations of light
reflected by exoplanets: 
neglecting polarization induces errors in numerically calculated fluxes
(thus also in e.g. the planet's albedo).
The reason for these errors is that light can only be fully
described by a 4-vector (see Sect.~\ref{stokesvectors}),
and a scattering process is only fully described by a 
$4 \times 4$ matrix. Consequently, the flux resulting from 
the scattering of unpolarized light differs usually from the 
flux resulting from the scattering of polarized light.
Because the unpolarized starlight that is incident on a planet
is usually polarized upon its first scattering, second and higher
orders of scattering induce errors in the fluxes when polarization
is neglected \citep[see also][]{1998GeoRL..25..135L,1994JQSRT..51..491M}.
For gaseous exoplanets, with their optically thick atmospheres,
the flux errors due to neglecting polarization can reach almost 
10~$\%$ \citep[][]{2005A&A...444..275S}. For Earth-like exoplanets,
with optically thinner atmospheres, we show in this paper 
that the errors are smaller: typically a few percent at short wavelengths
(~0.4~$\mu$m) and they decrease with wavelength
(see Sect.~\ref{section_summary}).

This paper has the following structure. 
In Sect.~\ref{sect_reflectedstarlight}, we describe 
how we define and calculate flux vectors and polarization for
extrasolar planets. 
In Sect.~\ref{section_modelplanets}, we describe the 
atmospheres and surfaces of our Earth--like model extrasolar planets.
In Sect.\ref{section_calculatedfluxes}, we present the numerically
calculated 
fluxes and degrees of polarization of starlight that is 
reflected by our Earth-like model planets 
for both horizontally
homogeneous planets and the so--called quasi horizontally
inhomogeneous planets, i.e. weighted mixtures of light 
reflected by horizontally homogeneous planets.
Section~\ref{section_summary}, finally, contains the summary
and discussion of our results.

%-----------------------------------------------------------------------
% Section 2
%-----------------------------------------------------------------------
\section{Describing and calculating reflected starlight}
\label{sect_reflectedstarlight}

%-----------------------------------------------------------------------
\subsection{Flux vectors and polarization}
\label{stokesvectors}

The flux (irradiance) and state of polarization of stellar light
that is reflected by a planet can fully be described by a flux 
(column) vector ${\bf F}$ as follows
\begin{equation}
   {\bf F}= 
   \left[ F, Q, U, V \right].
\label{eq_stokesvector}
\end{equation}
Here, $F$ is the total reflected flux divided, 
$Q$ and $U$ describe the linearly polarized flux, 
and $V$ the circularly polarized flux
\citep[see e.g.][]{1974SSRv...16..527H,2004Hovenier}. 
The fluxes $F$, $Q$, $U$, and $V$ in Eq.~\ref{eq_stokesvector} 
have the dimension W~m$^{-2}$~m$^{-1}$.

Fluxes $Q$ and $U$ are defined with respect to a
reference plane, for which we, unless stated otherwise,
chose the so-called planetary scattering plane, i.e. 
the plane through the centers of the
star and the planet, that also contains the observer.
We define $F_{\rm x^\circ}$ as the flux that is 
measured through a polarization filter oriented perpendicular to
the direction of propagation of the light, and with its
optical axis making an angle of $x^\circ$ with the reference plane.
The angle is measured rotating from the reference plane to 
the filter's optical axis in the anti--clockwise 
direction when looking in the direction of propagation of 
the light \citep[see][]{1974SSRv...16..527H,2004Hovenier}.
The so--called (linearly) polarized fluxes, $Q$ and $U$,
can then in principle be obtained with the following 
flux measurements 
\begin{eqnarray}
   Q & = & F_{0^\circ} - F_{90^\circ}, \label{eq_polmeas1} \\
   U & = & F_{45^\circ} - F_{135^\circ} \label{eq_polmeas2}.
\end{eqnarray}
Expressed in the fluxes of Eqs.~\ref{eq_polmeas1} 
and~\ref{eq_polmeas2}, the total flux, $F$, 
is simply equal to either $F_{0^\circ}+F_{90^\circ}$
or $F_{45^\circ}+F_{135^\circ}$.
Note that modern polarimetry has much more options available than
polarization filters, such as various types of
modulators \citep[see e.g.][]{2004A&A...422..703G,2004SPIE.5492..463G,
2005ASPC..343...89S,2006SPIE.6269E..26K}.

Flux vectors can be transformed from one reference plane
to another, e.g. from the planetary scattering plane
(which depends on the location of the planet on the sky
with respect to its star)
to the optical plane of a polarimeter, by multiplying
them with a so--called rotation matrix ${\bf L}$ that 
is given by (see Hovenier et al. 2004)
\begin{equation}
   {\bf L}(\beta)= \left[ \begin{array}{cccc}
      1 & 0 & 0 & 0 \\
      0 &  \cos{2\beta} & \sin{2\beta} & 0 \\
      0 & -\sin{2\beta} & \cos{2\beta} & 0 \\
      0 & 0 & 0 & 1
      \end{array} \right].
\label{eq_rotationmatrix}
\end{equation}
Angle $\beta$ is the angle between the two reference planes,
measured rotating in the anti--clockwise direction from the
old to the new plane when looking in the direction of
propagation of the light ($\beta~\geq~0$).

The direction of linear polarization with respect to the
reference plane is given by angle $\chi$,
which can be found using
\begin{equation}
   \tan 2\chi = U/Q,
\label{eq_anglepol}
\end{equation}
where the convention is to choose $\chi$ such that 
$0 \leq \chi < \pi$, and such that $\cos 2\chi$ 
and $Q$ have the same sign 
\citep[see][]{1974SSRv...16..527H,2004Hovenier}.
In particular, when $\chi=90^\circ$ ($\chi=0^\circ$), 
$Q < 0$ ($Q > 0$), and the direction of polarization
is perpendicular (parallel) to the reference plane,
i.e. perpendicular (parallel) to the imaginary line 
connecting the centers
of the star and the planet as seen from the observer.

The degree of polarization of the reflected starlight is
defined as
\begin{equation}
   P = \frac{\sqrt{Q^2 + U^2 + V^2}}{F}.
\label{eq_polarization}
\end{equation}
Note that $P$ as defined in Eq.~\ref{eq_polarization}
is independent of the choice of reference plane.
Assuming that the planet is mirror--symmetric with respect to
the planetary scattering plane, and assuming the incoming
starlight is unpolarized, the disk--integrated
Stokes parameters $U$ and $V$ will equal zero 
because of symmetry (the incoming starlight is unpolarized).
In that case, we can use an alternative definition of 
the degree of polarization
(see also Eq.~\ref{eq_polmeas1}), namely
\begin{equation}
   P_{\rm s} = - \frac{Q}{F} = 
  - \frac{F_{0^\circ}-F_{90^\circ}}{F_{0^\circ}+F_{90^\circ}}.
\label{eq_ps}
\end{equation}
For $P_{\rm s} > 0$, the light is polarized 
perpendicular to the reference plane (i.e. $\chi = 90^\circ$),
and for $P_{\rm s} > 0$, the light is polarized 
parallel to the reference plane (i.e. $\chi = 0^\circ$).
The subscript $s$ (from "sign") in $P_{\rm s}$ thus 
indicates that the 
direction of polarization is included in the definition of
the degree of polarization.

%-----------------------------------------------------------------------
\subsection{Computing reflected starlight}

\begin{figure}
\vspace*{5.0cm}
\centering
\resizebox{10cm}{!}{\includegraphics{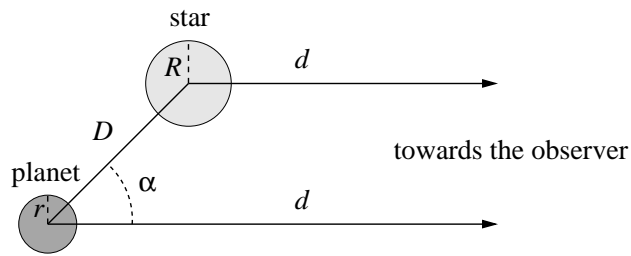}}
\caption{Top-view sketch of the geometries involved in observing 
         extrasolar planets: 
         $D$ is the distance between the star and the planet,
         $R$ is the radius of the star,
         $d$ is the distance between the planetary system 
         and the observer,
         $r$ is the radius of the planet, and
         $\alpha$ is the planetary phase angle.
         We assume that $D \gg R$ and that $d \gg r$.}
\label{fig1}
\end{figure}

Given a spatially unresolved spherical planet with radius $r$,
the flux vector ${\bf F}$ (see Eq.~\ref{eq_stokesvector})
of stellar light with wavelength $\lambda$ that
has been reflected by the planet and that arrives
at an observer at a distance $d$ (with $d \gg r$) can be written as 
\citep[see also][]{2006A&A...452..669S}
\begin{equation}
   {\bf F}(\lambda,\alpha) = \frac{r^2}{d^2} \hs 
                 \frac{1}{4} {\bf S}(\lambda,\alpha) \hs 
                 \pi {\bf F}_0(\lambda).
\label{eq_reflectedflux}
\end{equation}
Here, $\alpha$ is the planetary phase angle, i.e. 
the angle between the star and the observer as seen 
from the center of the planet.
Note that $\alpha = 180^\circ - \Theta$, with
$\Theta$ the total scattering angle of the incoming starlight.
A sketch of the geometries is given in Fig.~\ref{fig1}.

Furthermore in Eq.~\ref{eq_reflectedflux}, 
${\bf S}$ is the $4 \times 4$ planetary scattering matrix 
\citep[see below and][]{2006A&A...452..669S},
and ${\bf F}_0$ represents the flux (column) vector
describing the stellar light that is incident on the planet, 
with $\pi F_0$ the stellar flux that arrives at the planet 
(in W~m$^{-2}$~m$^{-1}$)
measured perpendicular to the direction of propagation of the
stellar light.
Integrated over the stellar disk, the stellar light of a solar--type
star can be assumed to be unpolarized \citep{1987Natur.326..270K}, 
hence in the following we will use 
${\bf F}_0(\lambda) = F_0(\lambda) [1,0,0,0] = F_0(\lambda) {\bf 1}$, 
with ${\bf 1}$ the unit column vector.
We assume that the starlight is unidirectional when it
arrives at the planet.

The planetary scattering matrix, ${\bf S}$, depends 
on the planetary phase angle, $\alpha$, and on the 
wavelength, $\lambda$. The relation between ${\bf S}$ 
and $\alpha$ and $\lambda$ depends on 
the composition and structure of the planetary atmosphere
and on the planetary surface. This dependence will be further 
described in Sect.~\ref{section_modelplanets}.
Using the planetary scattering plane as the reference plane,
and assuming the planet is mirror--symmetric with
respect to this reference plane, matrix ${\bf S}$ is 
given by \citep[see][]{2006A&A...452..669S,2004A&A...428..663S}
\begin{equation}
   {\bf S}(\lambda,\alpha)= \left[ \begin{array}{cccc}
      a_1(\lambda,\alpha) & b_1(\lambda,\alpha) & 0 & 0 \\
      b_1(\lambda,\alpha) & a_2(\lambda,\alpha) & 0 & 0 \\
      0 & 0 & a_3(\lambda,\alpha) & b_2(\lambda,\alpha) \\
      0 & 0 & -b_2(\lambda,\alpha) & a_4(\lambda,\alpha)
      \end{array} \right].
\label{eq_scatteringmatrix}
\end{equation}
Matrix element $a_1$ is usually called the 
{\em planetary phase function}.
Matrix ${\bf S}$ is normalized
such that the average of $a_1$ over all
directions equals the planet's (monochromatic)
{\em Bond} albedo, $A_{\rm B}$, which is the fraction of the
incident stellar flux that is reflected by
the planet in all directions, i.e.
\begin{equation}
   \frac{1}{4\pi} \int_{4\pi} a_1(\lambda,\alpha) \hs d\omega =
   \frac{1}{2} \int_{0}^{\pi} 
   a_1(\lambda,\alpha) \hs \sin \alpha \hs d\alpha
   \equiv A_{\rm B}(\lambda),
\label{eq_matrixnormalization}
\end{equation}
where $d \omega$ is an element of solid angle.
The (monochromatic) {\em geometric} albedo, $A_{\rm G}$, of a 
planet is the ratio of the flux reflected by the planet 
at $\alpha=0^\circ$, to the flux reflected by a 
Lambertian surface subtending the same solid angle 
(i.e. $\pi r^2 / d^2$) on the sky. Thus,
\begin{equation}
   A_{\rm G}(\lambda) = \frac{F(\lambda,0^\circ)}{F_0(\lambda)} \hs 
                  \hs \frac{d^2}{\pi r^2} 
                = \frac{1}{4}~a_1(\lambda,0^\circ).
\label{eq_geoalb}
\end{equation}
%Note that although an extrasolar planet will generally be at 
%its brightest at or close to $\alpha = 0^\circ$, it is then
%located behind its star, and thus impossible to observe.
%In addition, the chances for an Earth-like extrasolar planet to have
%a phase angle equal or close to $0^\circ$ 
%as seen from an observer in our Solar System 
%are very small (see Sect.~\ref{sect_yyy}).

From Eqs.~\ref{eq_reflectedflux} and~\ref{eq_scatteringmatrix}
it is clear that with unpolarized incident stellar light,
the observable total flux, $F$, of starlight that is 
reflected by a planet is given by
\begin{equation} 
   F(\lambda,\alpha)= \frac{r^2}{d^2} \hs
         \frac{1}{4} a_1(\lambda,\alpha) \hs
         \pi F_0(\lambda),
\label{eq_observableflux}
\end{equation}
and the observable polarized flux, $Q$, by
\begin{equation}
   Q(\lambda,\alpha)= \frac{r^2}{d^2} \hs
         \frac{1}{4} b_1(\lambda,\alpha) \hs
         \pi F_0(\lambda).
\label{eq_observableqlux}
\end{equation}
%Dividing this observable planetary flux to 
%the observable stellar flux gives the following expression
%for the relative planetary flux
%\begin{equation}
%   K(\lambda,\alpha)= \frac{r^2}{D^2} \hs \frac{1}{4} \hs
%   a_1(\lambda,\alpha).
%\label{eq_relativeflux}
%\end{equation}
%Eq.~\ref{eq_relativeflux} only holds for monochromatic
%fluxes, or for gray planets, i.e. with wavelength independent
%geometric albedo's.

With unpolarized incident stellar light {\em and} a planet
that is mirror-symmetric with respect to the planetary
scattering plane, it follows from Eqs.~\ref{eq_observableflux} 
and~\ref{eq_observableqlux} that 
the degree of polarization $P_{\rm s}$ of the starlight that 
is reflected by the planet can simply be rewritten as 
(cf. Eq.~\ref{eq_ps})
\begin{equation}
   P_{\rm s}(\lambda,\alpha) =
      - \frac{b_1(\lambda,\alpha)}{a_1(\lambda,\alpha)}. 
\label{eq_psb1}
\end{equation}
The degree of polarization of the reflected light then 
thus solely depends on the planetary scattering matrix 
elements $a_1$ and $b_1$. 
Because both $P$ (Eq.~\ref{eq_polarization}) and 
$P_{\rm s}$ (Eqs.~\ref{eq_ps} and~\ref{eq_psb1}) are relative measures,
they are independent of the radii $r$ and $R$,
the distances $D$ and $d$, which
is very convenient when analyzing
direct observations of extrasolar planets at unknown
distances.

To calculate the flux and degree of polarization 
of light reflected by a given Earth-like model planet 
(see Sect.~\ref{section_modelplanets})
across a given wavelength
region and for a given planetary phase angle,
we have to calculate elements of the planetary scattering 
matrix ${\bf S}$ (Eq.~\ref{eq_scatteringmatrix}). 
For this we use the algorithm as described
in \citet{2006A&A...452..669S}, 
which combines an accurate adding-doubling 
algorithm \citep{1980vandeHulst,1987A&A...183..371D}
to compute the radiative transfer through a locally 
plane-parallel planetary model atmosphere, and a fast, numerical, 
disk-integration algorithm, to 
integrate the reflected flux vectors 
across the illuminated and visible part of the planetary disk. 

Our disk-integration algorithm \citep[][]{2006A&A...452..669S}
is very efficient, and its computing time depends only 
little on the
number of planetary phase angles for which the disk-integrated
flux vectors are calculated.
The disadvantage of the current version of the algorithm is
that it can only handle horizontally homogeneous planets
(which are mirror-symmetric with respect to the 
reference plane).
Thus, while a planetary model atmosphere can be inhomogeneous
in the vertical direction, it varies neither with latitude
nor with longitude.
Calculated flux and polarization spectra of horizontally
homogeneous planets are presented in 
Sects.~\ref{sect_wavelengthindependent} 
and~\ref{sect_wavelengthdependent}.
In this paper, we will approximate horizontally inhomogeneous
planets by using weighted sums of horizontally homogeneous 
planets. The flux vector of such a {\em quasi} horizontally
inhomogeneous planet is calculated according to
\begin{equation}
   {\bf F}(\lambda,\alpha) = 
   \sum_{n=1}^{N} f_n {\bf F}_n(\lambda,\alpha)
   \hspace*{0.5cm} \mbox{with} \hspace*{0.2cm}
   \sum_{n=1}^{N} f_n = 1,
\label{eq_weightedsum}
\end{equation}
with $N$ the number of horizontally homogeneous planets.
Calculated flux and polarization spectra of quasi horizontally
inhomogeneous planets are presented in Sect.~\ref{sect_mixtures}.

%-----------------------------------------------------------------------
\subsection{Atmospheric extinction and instrumental response}
\label{section_atmosphereinstrument}

The flux vector ${\bf F}$ as described in 
Eq.~\ref{eq_reflectedflux}
includes neither extinction in the terrestrial atmosphere,
nor the response of an instrument. It thus pertains to the flux
vector as it can be observed in space.
Adding atmospheric extinction and/or 
instrumental effects is 
straightforward by multiplying vector ${\bf F}$
from Eq.~\ref{eq_reflectedflux} with the matrix describing
atmospheric extinction, and/or with the matrix describing 
the instrumental response.

Atmospheric extinction usually affects only the flux of the
directly transmitted light, not its state of polarization,
and can then simply be described by the scalar
$\exp({-\tau}) \cos^{-1}(\theta_0)$, with  
$\tau$ the (wavelength dependent) extinction optical thickness 
of the atmosphere between the observer and space, and 
$\theta_0$ the zenith angle of the observed exoplanet.

Since most instruments not only affect the flux of the light that
enters the instrument, but also its state of polarization,
\citep[see e.g.][regarding polarization sensitive Earth-observation
instruments]{2000JGR...10522379S}
an instrumental response matrix can be quite complicated. 
With a polarization sensitive instrument, the flux that is
measured will not only depend on the flux
$F$ (see Eq.~\ref{eq_observableflux}) 
of the light that is reflected by the planet,
but also on its state of polarization.
Of course, a polarization sensitive instrument will usually 
also change the state of polarization of the observed light.
Thus, when analyzing flux and/or polarization observations 
of starlight that is reflected by an exoplanet, one has to
properly account for the polarization sensitivity of 
one's instrument.

In this paper, we will ignore both atmospheric extinction 
and instrumental effects (apart from the spectral resolution
of 0.001~$\mu$m), and thus limit ourselves to the 
total flux and degree of polarization as they can be observed
in space. Since the calculated total and
polarized fluxes will be made available through the CDS, 
the atmospheric extinction and/or instrumental 
effects that particular observations require can be applied.

%-----------------------------------------------------------------------
\section{The model planets}
\label{section_modelplanets}

The atmospheres of our Earth-like model planets are described by 
stacks of homogeneous layers containing gaseous molecules and, 
optionally, cloud particles.  
Each model atmosphere is bounded below by a flat, 
homogeneous surface.
In the next subsections, we will describe the composition,
structure, and optical properties of our model atmospheres
(Sect.~\ref{section_modelatmospheres}),
and the reflection properties of our model surfaces
(Sect.~\ref{section_modelsurfaces}).

%-----------------------------------------------------------------------
\subsection{The model atmospheres}
\label{section_modelatmospheres}

All of our model atmospheres consist of 16 homogeneous layers.
For the radiative transfer calculations, we need to know for each
atmospheric layer: its optical thickness, $b$,
and the single scattering albedo, $a$, and scattering 
matrix, ${\bf F}_{\rm sca}$, (see Hovenier et al., 2004) 
of the mixture of molecules and cloud particles.

An atmospheric layer's optical thickness, $b$, is the sum of
its molecular and cloud extinction optical thicknesses,
$b^{\rm m}$ and $b^{\rm c}$, i.e.
\begin{equation}
   b(\lambda)= b^{\rm m}(\lambda)+b^{\rm c}(\lambda)
              = b^{\rm m}_{\rm sca}(\lambda)+ 
                b^{\rm m}_{\rm abs}(\lambda)+
                b^{\rm c}_{\rm sca}(\lambda)+
                b^{\rm c}_{\rm abs}(\lambda). 
\label{eq_blayer}
\end{equation}
Here, $b^{\rm m}_{\rm sca}$ and $b^{\rm m}_{\rm abs}$ are 
the molecular scattering and absorption optical
thicknesses, respectively, and $b^{\rm c}_{\rm sca}$ and 
$b^{\rm c}_{\rm abs}$ are the cloud scattering and
absorption optical thicknesses.

The molecular scattering optical thickness 
of each atmospheric layer, $b^{\rm m}_{\rm sca}$,
is calculated as described in 
\citet{2000JQSRTStam}, and depends a.o. on the molecular 
column density (molecules per m$^{2}$), the 
refractive index of dry air under
standard conditions \citep{1972JOSA...62..958P}, 
and the depolarization factor of air, 
for which we adopt the (wavelength dependent) values 
provided by \citet{1984P&SS...32..785B}.
The molecular column density depends on the ambient pressure and
temperature, the vertical profile of which is given in
Table~\ref{tab1} \citep{1972McClatchey} (to avoid introducing
too many variables, we use this 
mid-latitude summer vertical profile for each model atmosphere).
In Table~\ref{tab1}, we also give for each atmospheric layer
$b^{\rm m}_{\rm sca}$ as we calculated at $\lambda=0.55~\mu$m.

The molecular absorption optical thickness
of each atmospheric layer, $b^{\rm m}_{\rm abs}$, depends on 
the molecular column density, the mixing ratios of the
absorbing gases, and their molecular absorption cross-section 
(in m$^2$ per molecule) 
\citep[see][for the details]{1999JGR...10416843S,
2000JQSRTStam}.
The terrestrial atmosphere contains numerous types of absorbing gases. 
In the wavelength region of our interest, i.e. between
0.3~$\mu$m and 1.0~$\mu$m, the main gaseous absorbers (and the only
absorbers we take into account here) are ozone (O$_3$),
oxygen (O$_2$) and water (H$_2$O). 
Unless stated otherwise, the mixing ratio of O$_2$ is
2.1$\cdot$10$^{4}$ ppm (parts per million) throughout each model
atmosphere. The altitude dependent mixing ratios of the trace
gases O$_3$ and H$_2$O are given in Table~\ref{tab1}.
We calculate the molecular absorption cross--sections of O$_2$,
O$_3$, and H$_2$O using absorption line data from 
\citet{2005JQSRT..96..139R}.
Because the absorption cross--sections of O$_2$ and H$_2$O are
rapidly varying functions of the wavelength, we have transformed
them into so--called $k$-distributions 
\citep[see][]{1991JGR....96.9027L,2000JQSRTStam}, 
using a wavelength spacing of 0.001~$\mu$m, a spectral
resolution of 0.001~$\mu$m, 20~Gaussian abscissae per
wavelength interval of 0.001~$\mu$m, and a block-shaped 
instrumental response function.
The absorption cross--sections of O$_3$ vary only
gradually with wavelength (at least 
between 0.3~$\mu$m and 1.0~$\mu$m); we assume them to be
constant across each wavelength interval of 0.001~$\mu$m.
Molecular absorption cross--sections in general not only depend
on the wavelength, but also on the ambient pressure and temperature.
For the purpose of this paper, i.e. presenting flux and polarization
spectra of 
Earth-like extrasolar planets and addressing the occurence of
spectral features in them, we use the absorption cross--sections
calculated for the lowest atmospheric layer (see Table~\ref{tab1}) 
throughout our model atmospheres. 

For each wavelength and each atmospheric layer, 
the cloud scattering and absorption optical thicknesses,
$b^{\rm c}_{\rm sca}$ and $b^{\rm c}_{\rm abs}$, are
calculated from the user-defined cloud particle column density
(in cloud particles per m$^{2}$), and the extinction 
cross--section and the single scattering 
albedo of the cloud particles.
The only cloud particles we will consider in this paper are spherical,
homogeneous, watercloud droplets. These droplets are distributed 
in size according to the standard size distribution described
by \citet{1974SSRv...16..527H},
with an effective radius of 2.0~$\mu$m and an effective variance
of 0.1. The refractive index is chosen to be wavelength independent
and equal to $1.33+0.0001$.
We calculate the extinction cross--section, single scattering albedo, 
and the scattering matrix, ${\bf F}^{\rm c}_{\rm sca}$, 
of the cloud droplets for wavelengths between
0.3~$\mu$m and 1.0~$\mu$m
using Mie-theory \citep[see][]{1957vandeHulst,1984A&A...131..237D}.
Unless stated otherwise, we assume a cloud with an optical thickness,
$b^a$, of~10 at $\lambda=0.55~\mu$m, with its bottom at 802~hPa
and its top at 628~hPa
(according to Table~\ref{tab1}, it thus extends vertically from 2~km 
to 4~km).
Because of the wavelength dependence of the droplets' 
extinction cross-section, the cloud's optical thickness 
varies with wavelength. In particular, at $\lambda=0.3~\mu$m,
$b^{\rm c}=9.6$, and at $\lambda=1.0~\mu$m, $b^{\rm c}=10.7$.

The single scattering albedo of the mixture of gaseous molecules
and cloud particles in an atmospheric layer is calculated 
according to
\begin{equation}
   a(\lambda) = \frac{b^{\rm m}_{\rm sca}(\lambda) + 
                      b^{\rm c}_{\rm sca}(\lambda)}
                     {b^{\rm m}_{\rm sca}(\lambda) + 
                      b^{\rm m}_{\rm abs}(\lambda) + 
                      b^{\rm c}_{\rm sca}(\lambda) + 
                      b^{\rm c}_{\rm abs}(\lambda)},
\end{equation}
and the scattering matrix \citep[see][]{2004Hovenier} 
of the mixture according to
\begin{equation}
   {\bf F}_{\rm sca}(\Theta,\lambda) = 
             \frac{b^{\rm m}_{\rm sca}(\lambda) 
                  {\bf F}^{\rm m}_{\rm sca}(\Theta,\lambda) + 
                  b^{\rm c}_{\rm sca}(\lambda) 
                  {\bf F}^{\rm c}_{\rm sca}(\Theta,\lambda)}
                  {b^{\rm m}_{\rm sca}(\lambda) + 
                  b^{\rm c}_{\rm sca}(\lambda)},
\end{equation}
where $\Theta$ is the scattering angle (with $\Theta=0^\circ$ 
indicating forward scattering), and
${\bf F}^{\rm m}_{\rm sca}$ and ${\bf F}^{\rm c}_{\rm sca}$ are 
the scattering matrices of, respectively, the molecules and
the cloud particles.
The scattering matrix ${\bf F}^{\rm m}_{\rm sca}$ of the gaseous 
molecules is calculated as described by 
\citet{2002JGRD.107t.AAC1S}, using the (wavelength dependent) 
depolarization factor of air \citep{1984P&SS...32..785B}. 
We do not explicitly account for
rotational Raman scattering, an inelastic molecular scattering process
\citep[see e.g.][]{1962Natur.193..762W,2001GeoRL..28..519A,2002JGRD.107t.AAC1S,2005JQSRT..95..309V,2005Icar..173..254S}, which gives rise to
a slight "filling-in" of high-spectral resolution features in 
reflected light spectra, such as stellar Fraunhofer lines
and gaseous absorption bands.
Each scattering matrix is normalized such that the average of the 
phase function, which is represented by scattering 
matrix element $F^{\rm 11}_{\rm sca}$, 
over all scattering directions is one
\citep[see][]{1974SSRv...16..527H,2004Hovenier}.

\begin{figure}
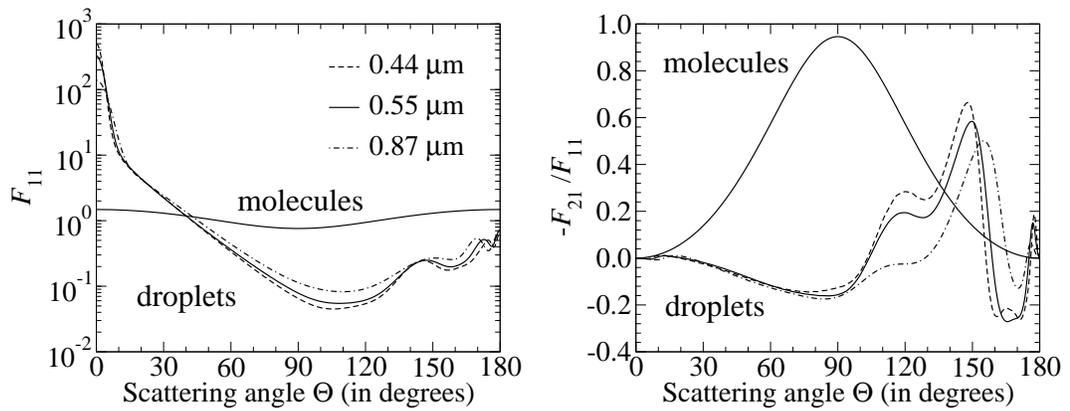

\vspace*{5.0cm}
\centering
\resizebox{14cm}{!}{\includegraphics{fig3a.eps} \hspace*{1.0cm}
                    \includegraphics{fig3b.eps} }
\vspace*{0.5cm}
\caption{The phase function (scattering matrix element 
         $F_{\rm 11}$) (on the left) and 
         the degree of linear polarization
         (-$F_{\rm 21}/F_{\rm 11}$) (on the right) 
         as functions of the scattering
         angle $\Theta$, for light singly scattered by gaseous 
         molecules and cloud droplets. The wavelength of the light is
         0.44~$\mu$m (dashed line, only for the cloud droplets), 
         0.55~$\mu$m (solid lines), and
         0.87~$\mu$m (dotted line, only for the cloud droplets).}
\label{fig3}
\end{figure}

Figure~\ref{fig3}a shows the phase functions of the gaseous
molecules and the cloud droplets at $\lambda=0.55~\mu$m.
To illustrate the wavelength dependence of the elements
of the cloud droplets' scattering matrix, we have also
plotted curves for $\lambda=0.44~\mu$m and $\lambda=0.87~\mu$m 
(these particular wavelengths will be used again later on,
in Sect.~\ref{section_calculatedfluxes}).
For the same wavelengths, Fig.~\ref{fig3}b shows the degree 
of linear polarization, $P_{\rm s}$ (Eq.~\ref{eq_ps}), 
of light that is singly scattered 
by the molecules and the cloud droplets as functions of the
single scattering angle, $\Theta$, assuming unpolarized
incident light.
The reference plane for this singly scattered light is the
plane through the incoming and the scattered light beams.
Note that the phase function and degree of polarization
of light singly scattered by gaseous molecules depends
on the wavelength, too, through the wavelength dependence
of the depolarization factor
\citep[see][]{1984P&SS...32..785B}, but only slightly so.

From Fig.~\ref{fig3}, it is clear that both the phase function 
and the degree of polarization pertaining 
to single scattering by molecules vary smoothly with the 
single scattering angle $\Theta$. 
The degree of polarization, $P_{\rm s}$, of the light that is
singly scattered by the molecules is positive 
(i.e. the direction of polarization is 
perpendicular to the reference plane) for all values of $\Theta$.
Furthermore, $P_{\rm s}$ of this light is highest at 
$\Theta=90^\circ$. At this scattering angle,
the light is not completely (i.e. 100~$\%$) polarized, but "only"
about 95~$\%$, because of the molecular depolarization factor
\citep{1984P&SS...32..785B}.
Both for the light scattered by the molecules
and for the light scattered by the cloud droplets,
$P_{\rm s}$ vanishes for $\Theta=0^\circ$ and 
$\Theta=180^\circ$, because of symmetry.

The degree of polarization of the light scattered by the cloud
droplets (Fig.~\ref{fig3}b) changes sign (i.e. the direction of 
polarization changes with respect to the reference plane)
a number of times between
$\Theta=0^\circ$ and $\Theta=180^\circ$, and 
shows strong angular features, in particular in the
backward scattering directions ($\Theta > 90^\circ$).
The peak in the polarization occuring
at $\Theta=148^\circ$ for $\lambda=0.44~\mu$m, 
at $\Theta=150^\circ$ for $\lambda=0.55~\mu$m, and 
at $\Theta=155^\circ$ for $\lambda=0.87~\mu$m, pertains to
what is commonly known as the primary rainbow, which is 
due to light that has been reflected inside the droplets
once.
The primary rainbow is seen in the flux phase functions 
(Fig.~\ref{fig3}a), too, only less prominent than in $P_{\rm s}$.
The angular features in $P_{\rm s}$ near $\Theta=120^\circ$ 
pertain to the secondary rainbow, which is due to light
that has been reflected inside the the droplets twice.
In the cloud droplets' phase functions (Fig.~\ref{fig3}a), 
only a hint of the secondary rainbow can be seen 
and only for $\lambda=0.44~\mu$m.
The occurence of a rainbow in reflected light is a strong 
and well-known
indicator for spherically shaped atmospheric particles, 
see e.g. \citet{1974SSRv...16..527H}, 
and more recently \citet{2002GeoRL..29i..27L} and references therein,
and \citet{2007AsBio...7..320B}.

%-----------------------------------------------------------------------
\subsection{The model surfaces}
\label{section_modelsurfaces}

To describe the reflection of light by the homogeneous, 
locally flat surfaces below the atmospheres of our model planets, 
we have to specify the surface reflection matrix, ${\bf A}_{\rm s}$.
The surface reflection matrix is normalized such that the
average of matrix element (1,1) of ${\bf A}_{\rm s}$ over all 
reflection directions equals the surface albedo, 
i.e. the fraction of the incident stellar flux that the surface 
reflects in all directions.
We'll denote the surface albedo by $A_{\rm s}$.
%For an incoming flux vector, multiplication with 
%the surface reflection matrix ${\bf A}^{\rm s}$ provides the 
%reflected flux vector in any direction. 

The Earth's surface is covered by numerous surface types
with myriads of (wavelength dependent) albedos, many of
which vary with e.g. their moistness and/or the season.
To avoid making our model planets too detailed at this stage,
we will compose the surfaces of our Earth-like model planets
out of only two surface types: (deep) ocean and (green) vegetation.
We assume that the surface that is covered by vegetation
completely depolarizes all incident light, i.e. except for element
element (1,1) of ${\bf A}_{\rm s}$, all elements of
the vegetation's surface reflection matrix equal zero. 
In addition, we assume that the reflection by the surface is
isotropic, i.e. reflection matrix element (1,1)
is independent of the directions of both the incoming and the
reflected light; element (1,1) thus simply equals the 
surface albedo, $A_{\rm s}$.
We thus describe a surface that is covered with vegetation
as a Lambertian reflecting surface.
In future studies, it will be interesting to include polarizing 
effects of vegetation, as presented by \citet{2007AAS...210.0906W}.

In Fig.~\ref{fig4}, we have plotted measured, wavelength dependent 
albedos of three types of vegetation: conifers, deciduous forest, 
and grass
\footnote{These three albedos have been taken from the ASTER Spectral
Library through the courtesy of the Jet Propulsion Laboratory,
California Institute of Technology, Pasadena, California.}
These albedo spectra share the following characteristics:
(1)~a local maximum between
0.5~$\mu$m and 0.6~$\mu$m, that is mainly due to the presence of
two absorption bands of chlorophyll, one near 0.45~$\mu$m and 
one near 0.67~$\mu$m, and (2)~a high albedo at
wavelengths longer than about 0.7~$\mu$m, that is related to 
the internal leaf and cell structure.
The sudden increase of the surface albedo at wavelengths
longer than 0.7~$\mu$m, is usually referred to as the 
{\em red edge} \citep[for an elaborate description of the
red edge, see][]{2005AsBio...5..372S}.
The slight decrease in the albedo around 0.97~$\mu$m is due
to absorption by water in the leaves. Stronger absorption 
bands of water occur at wavelengths longer than 1.2~$\mu$m.
Because in this paper we do not study the effects of
differences in the albedos of different 
types of vegetation on the light that is reflected by a planet,
we will only use the wavelength dependent 
albedo of the deciduous forest to represent the reflectivity 
by vegetation on our model planets.

Whether vegetation on Earth-like extrasolar planets will have
the same spectral features, in particular the red edge, 
as we find on Earth, is still an open question 
\citep[see][]{2002Icar..157..535W}.
Model studies for albedos of vegetation on Earth-like planets 
around M stars have been published by
\citet{2007AsBio...7..252K,
2005AsBio...5..706S,2006ApJ...644L.129T}.
For our purpose, presenting flux and polarization spectra
and in particular their differences and similarities without
focussing on the detection of features, an 
Earth-like vegetation albedo is sufficient.

Although on Earth, deep oceans do show some color, especially
in shallow regions where algae and other small organisms bloom, 
for the purpose of this paper it is safe to simply 
assume the oceans are black across the wavelength interval of 
our interest, i.e. from 0.3 to 1.0~$\mu$m.
Even with an albedo $A_{\rm s}$ equal to zero, however, 
our model oceans do reflect a fraction of the
light that is incident on them,
because we include a specular (i.e. Fresnel) 
reflecting interface between the atmosphere and the black ocean.
Specular reflection is anisotropic and generally leads to  
polarized reflected light. 
We use the specular reflection matrix as described 
by \citet{1997Haferman}, with a (wavelength independent) 
index of refraction that is equal to 1.34.
Our model ocean surface is flat, i.e. there are no
waves. The influence of oceanic waves
will be subject of later studies,
using the wave distribution model by \citet{1954JOSA...44..838C}
\citep[for a recent evaluation of this model, see][]{2006JGRC..11106005B},
which can be included in our adding-doubling radiative transfer model 
\citep[see e.g.][]{2002JAtS...59..383C}.
We neglect the contribution of whitecaps
to the ocean albedo, which appears to be a valid assumption for
average wind speeds measured on the Earth's oceans
\citep[][]{1984ApOpt..23.1816K}.

%------------------------------------------------------------------
% Section 3
%------------------------------------------------------------------
\section{Calculated flux and polarization spectra}
\label{section_calculatedfluxes}

In this section, we will present the numerically calculated
total flux and degree of polarization of starlight that is 
reflected by Earth-like model planets as described in the 
previous section (Sect.~\ref{section_modelplanets}).
The reflected flux, $F$, is calculated according to 
Eq.~\ref{eq_reflectedflux}. Unless stated otherwise, 
we assume that $r=1$, $d=1$, and $\pi F_{\rm 0}=1$,
independent of $\lambda$.
With unpolarized incident light, $F$
thus equals $\frac{1}{4} a_{\rm 1}$, which
is the planet's geometric albedo $A_{\rm G}$ 
in case planetary phase angle $\alpha=0^\circ$ 
(see Eq.~\ref{eq_geoalb}).
The degree of polarization is calculated according
to Eq.~\ref{eq_psb1}, and thus includes the direction of 
polarization. 

Tables containing elements $a_1$ and $b_1$ of the planetary
scattering matrix ${\bf S}$ as functions of the wavelength,
and as functions of the planetary phase angle, for the various
horizontally homogeneous model planets that are presented
in the following sections will be made available through
the CDS.
From the elements $a_1$ and $b_1$, and given distance $d$,
planetary radius $r$, and the (wavelength dependent) 
incident stellar flux (e.g. in W m$^{-2}$ m$^{-1}$), 
the observable total flux, 
$F$, polarized flux, $Q$, and degree of
polarization, $P_{\rm s}$, can be calculated using 
Eqs.~\ref{eq_observableflux},~\ref{eq_observableqlux},
and~\ref{eq_psb1}, respectively.

%------------------------------------------------------------------
\subsection{Clear planets with wavelength independent surface 
            albedos} 
\label{sect_wavelengthindependent}

%------------------------------------------------------------------
\subsubsection{Wavelength dependence}

Figure~\ref{fig5} shows the wavelength dependence of 
the total flux, $F$, and the degree of polarization, $P_{\rm s}$, 
of starlight that is reflected by 
six~Earth--like model planets with similar, clear (i.e. cloudless) 
atmospheres, and Lambertian reflecting (i.e. isotropically 
reflecting and completely depolarizing) surfaces with 
wavelength independent albedos, $A_{\rm s}$, ranging from 0.0 
to 1.0. 
The planetary phase angle, $\alpha$, is 90$^\circ$, i.e. half
of the observable planetary disk is illuminated by the star.
As explained in Sect.~\ref{sect_introduction}, the probability
to directly observe an exoplanet at or near this phase angle
(quadrature) is relatively high (provided there is an observable 
exoplanet). 

Each curve in Fig.~\ref{fig5} can be thought of to consist
of a continuum with superimposed high--spectral resolution 
features. 
The continua of the flux and polarization curves are
determined by the scattering of light by gaseous 
molecules in the atmosphere and by the surface albedo. 
The high--spectral resolution features are due to the
absorption of light by the gases O$_3$, O$_2$, and H$_2$O
(see below).
Note that strength and shape of the absorption bands 
depend on the spectral resolution (0.001~$\mu$m) of 
the numerical calculations.

In the total flux curves (Fig.~\ref{fig5}a),
the contribution of light scattered by atmospheric molecules 
is largest around 0.34~$\mu$m; 
at shorter wavelengths, light is absorbed by O$_3$
in the so-called Huggins absorption band, and at longer 
wavelengths, the amount of starlight that is scattered by the
atmospheric molecules decreases, simply because the atmospheric
molecular scattering optical thickness decreases with wavelength,
as $b^{\rm m}_{\rm sca}$ is roughly proportional to 
$\lambda^{-4}$ \citep[see e.g.][]{2000JQSRTStam}.
For the planet with the black surface ($A_{\rm s}=0.0$), where 
the only light that is reflected by the planet comes from 
scattering by atmospheric molecules, the
flux of reflected starlight decreases towards zero
with increasing wavelength. For the planets with reflecting 
surfaces, the contribution of light that is reflected by 
the surface to the total reflected flux increases
with increasing wavelength.
Because the surface albedos are wavelength independent, the 
continua of the reflected fluxes become wavelength independent,
too, at the longest wavelengths. This is not obvious from 
Fig.~\ref{fig5}a, because of the presence of 
high-spectral resolution features.

The high--spectral resolution features in the flux curves 
of Fig.~\ref{fig5} are all due to gaseous absorption bands.
As mentioned above, at the shortest wavelengths, 
light is absorbed by~O$_3$.
The so--called Chappuis absorption band of O$_3$ gives
a shallow depression in the flux curves that is 
visible between about 0.5~$\mu$m and 0.7~$\mu$m, in particular
in the curves pertaining to a high surface albedo. 
The flux curves contain four absorption bands of O$_2$, 
i.e. the $\gamma$-band around 0.63~$\mu$m, 
the $B$-band around 0.69~$\mu$m, 
the conspicuous $A$-band around 0.76~$\mu$m, 
and a weak band around 0.86~$\mu$m.
These absorption bands, except the $A$-band, are
difficult to identify from Fig.~\ref{fig5}a,
because they are located either next to or within one of the
many absorption bands of H$_2$O (which are all the 
bands not mentioned previously).

The polarization curves (Fig.~\ref{fig5}b) are, like the 
flux curves, shaped by light scattering and absorption 
by atmospheric molecules, and by the surface reflection.
The contribution of the scattering by atmospheric molecules
is most obvious for the planet with the black surface 
($A_{\rm s}=0.0$),
where there is no contribution of the surface to the 
reflected light.
For this model planet and phase angle, 
$P_{\rm s}$ has a local minimum around 0.32~$\mu$m. 
At shorter wavelengths, $P_{\rm s}$ is
relatively high because there the absorption of light 
in the Huggins band of O$_3$ decreases the amount 
of multiple scattered
light, which usually has a lower degree of polarization
than the singly scattered light.
In general, with increasing atmospheric absorption optical
thickness, $P_{\rm s}$ will tend towards the degree of 
polarization
of light singly scattered by the atmospheric constituents
(for these model planets: only gaseous molecules), which
depends strongly on the single scattering angle $\Theta$ 
and thus on the planetary phase angle $\alpha$. 
From Fig.~\ref{fig3}b, it can be seen that at a scattering 
angle of 90$^\circ$, $P_{\rm s}$ of 
light singly scattered by gaseous molecules is about 0.95.
This explains the high values of $P_{\rm s}$ at the shortest
wavelengths in Fig.~\ref{fig5}b. 
With increasing wavelength, the amount
of multiple scattered light decreases, simply because of
the decrease of the atmospheric molecular scattering optical
thickness. Consequently, $P_{\rm s}$ 
of the planet with the black surface increases with wavelength, 
to approach its single scattering value at the
smallest scattering optical thicknesses.

With a reflecting surface below 
the atmosphere, $P_{\rm s}$ also tends to its single scattering 
value at the shortest wavelengths, because with increasing
atmospheric absorption optical thickness, the contribution of
photons that have been reflected by the depolarizing 
surface to the total number of reflected photons
decreases (both because with absorption
in the atmosphere, less photons reach
the surface and less photons that have been reflected 
by the surface reach the top of the atmosphere)
\citep[see e.g.][]{1999JGR...10416843S}.
In case the planetary surface is reflecting, $P_{\rm s}$
of the planet will start to decrease with wavelength,
as soon as the contribution of photons that have
been reflected by the depolarizing surface to the 
total number of reflected photons becomes significant. 
As can been seen in Fig.~\ref{fig5}b, 
the wavelength at which the decrease of $P_{\rm s}$ starts 
depends on the surface albedo: the higher the albedo,
the shorter this wavelength. 
It is also obvious that with increasing wavelength,
the sensitivity of $P_{\rm s}$ to $A_{\rm s}$ decreases.
This sensitivity clearly depends on the atmospheric
molecular scattering optical thickness. 

Like with the flux curves, the high--spectral resolution features
in the polarization curves of Fig.~\ref{fig5}b are all due
to gaseous absorption. 
The explanation for the increased degree of polarization inside
the O$_2$ and H$_2$O absorption bands is the same as that given
above for the Huggins absorption band of O$_3$:
with increasing atmospheric absorption optical
thickness, the contribution of multiple scattered light 
to the reflected light decreases, and hence $P_{\rm s}$ increases
towards the degree of polarization of light singly scattered
by the atmospheric constituents, i.e. gaseous molecules.
In case atmospheres contain aerosol and/or cloud
particles, $P_{\rm s}$ both inside and outside the absorption
bands will depend on the single scattering properties of 
those aerosol and/or cloud particles, too
\citep[see][for a detailed description of $P_{\rm s}$ across
gaseous absorption lines]{1999JGR...10416843S}.
\citet{2004A&A...428..663S} and \citet{2003toed.conf..615S}
show calculated polarization spectra of Jupiter-like
extrasolar planets with gaseous absorption bands due to 
methane.
An increase of the degree of polarization across gaseous
asorption bands has been measured in so--called zenith
sky observations on Earth 
\citep[][]{1994STAMMES,1995PREUSKER,1997ABEN,1999GeoRL..26..591A}, 
and, recently, in observations of Jupiter, Uranus and Neptune 
\citep[][]{2007A&A...463.1201J,2006A&A...452..657S,
2005ASPC..343..189J}, with
methane as the absorbing gas. 

It is interesting to note that the polarization
spectrum of an extrasolar planet will generally be 
{\em in}sensitive to 
absorption that takes place between the planet and the observer
because it is a relative measure 
(see Eqs.~\ref{eq_polarization} and~\ref{eq_ps}).
Thus, if the telescope were located on the Earth's surface,
the polarization features as shown in Fig.~\ref{fig5}b 
would be unaffected by absorption within the Earth's atmosphere;
polarimetry would in principle allow the detection of e.g. 
O$_2$ in an 
extrasolar planetary atmosphere despite the O$_2$ in the Earth's 
atmosphere (the number of photons received by the telescope, i.e.
the flux, would of course be strongly affected by absorption in the
Earth's atmosphere).

%-------------------------------------------------------------------
\subsubsection{Phase angle dependence}
\label{section_phaseangle}

Figure~\ref{fig6} shows the phase angle dependence of 
the total flux, $F$, and the degree of polarization, $P_{\rm s}$, 
of the starlight 
that is reflected by three of the six Earth-like planets appearing in
the previous section, namely, the planets with $A_{\rm s}=0.0$,
0.4, and 1.0. The phase angle dependence has been plotted
for two wavelengths: 0.44~$\mu$m and 0.87~$\mu$m.
Remember that the flux at 
phase angle $\alpha=0^\circ$ is just the planet's geometric albedo 
$A_{\rm G}$ (Eq.~\ref{eq_geoalb}). 
For $A_{\rm s}=0.0$, $A_{\rm G}=0.14$ at $\lambda=0.44~\mu$m,
and $A_{\rm G}=0.011$ at $\lambda=0.87~\mu$m.
For $A_{\rm s}=0.4$ and 1.0, we find, respectively,
$A_{\rm G}=0.34$ (0.44~$\mu$m) and 0.27 (0.87~$\mu$m), 
and 0.72 (0.44~$\mu$m) and 0.67 (0.87~$\mu$m)
(see Fig.~\ref{fig6}a).
The planets' geometric albedos at $\lambda=0.87~\mu$m
are close to $\frac{2}{3} \hs A_{\rm s}$, i.e.\
the geometric albedo of a planet with a Lambertian 
reflecting surface but without an atmosphere
\citep[see][]{2006A&A...452..669S}, because at
this wavelength, the scattering
optical thickness of the model atmosphere is only 0.015.
At 0.44~$\mu$m, this optical thickness is 0.24,
and the light scattered within the model atmosphere
does contribute significantly to the planet's geometric
albedo, especially when $A_{\rm s}$ is small.

The strong phase angle dependence of $F$ (Fig.~\ref{fig6}a) is, 
for a given value of $A_{\rm s}$,
largely due to the variation of the illuminated and visible fraction
of the planetary disk with the phase angle.
Other variations are related to the reflection properties of the 
surface and the scattering properties of the overlying atmosphere.
These variations can be seen more clearly in Fig.~\ref{fig7},
where we show the curves of Fig.~\ref{fig6}a normalized
at $\alpha=0^\circ$.
The curves for $\lambda=0.87~\mu$m and $A_{\rm s}=0.4$ and
$A_{\rm s}=1.0$ coincide with the theoretical (normalized) curve
expected for Lambertian reflecting spheres
\citep[see][]{1980vandeHulst,2006A&A...452..669S}, 
because, as explained above, at this wavelength, 
the contribution of light scattered within the model 
atmosphere is almost negligible.

In Fig.~\ref{fig6}b, we show the degree of polarization, 
$P_{\rm s}$, as a function of the 
phase angle. Like the reflected flux, $P_{\rm s}$ depends 
strongly on the phase angle.
Note that for $\alpha=0^\circ$ and $\alpha=180^\circ$,
$P_{\rm s}$ equals zero because of symmetry (the incoming 
starlight is unpolarized).
For the planet with the black surface, $P_{\rm s}$ appears to
be fairly symmetric around $\alpha=90^\circ$, mainly because
the degree of polarization of light singly scattered by
gaseous molecules is symmetric around $\Theta=90^\circ$
(see Fig.~\ref{fig3}b). This symmetry is particularly 
apparent for \mbox{$\lambda=0.87~\mu$m}, where there is much less 
multiple scattering than for $\lambda=0.44~\mu$m. 
Across most of the phase angle range, $P_{\rm s}$ of the 
planet with the black surface is positive,
indicating that the reflected light is polarized perpendicular to the
scattering plane (i.e. perpendicular to the imaginary
line connecting the planet and the star as seen by the observer). 
Only for the largest scattering angles, $P_{\rm s}~<~0$
(which is difficult to see in Fig.~\ref{fig6}b 
for $\lambda=0.87~\mu$m).
At these angles, the reflected light is thus polarized parallel 
to the scattering plane (i.e. parallel to the imaginary line 
connecting the planet and the star).
The negative polarization is mainly due to second order
scattered light: although this light comprises only a small fraction
of the reflected light at these large phase angles, it does
leave its traces in the polarization signature of the planet, 
because the degree of polarization of the first order 
scattered light (which is the main contributor to the reflected
light) is close to zero. 
For $\lambda=0.44~\mu$m, $P_{\rm s}<0$ when $\alpha>164^\circ$,
while for $\lambda=0.87~\mu$m, $P_{\rm s}<0$ only when 
$\alpha>174^\circ$, because with increasing wavelength, the
amount of second order scattered 
light decreases, and with that the phase angle at which the
second order scattered light changes the sign of $P_{\rm s}$
increases.

For the planets with the reflecting surfaces, the maximum
degree of polarization occurs at phase angles larger than 90$^\circ$
(see Fig.~\ref{fig6}b). In particular, with increasing wavelength
and/or increasing surface albedo, the maximum degree of polarization
shifts to larger phase angles, because with increasing $\alpha$,
the fraction of reflected light that 
has touched the depolarizing surface at least once, decreases.
The contribution of light that has been polarized within 
the planetary atmosphere to the reflected light thus increases
with increasing $\alpha$.
This also explains why at the largest values of $\alpha$, 
$P_{\rm s}$ in Fig.~\ref{fig6}b, is independent of $A_{\rm s}$.
%The longer the wavelength, and thus the smaller the
%atmospheric scattering optical thickness, the larger $\alpha$
%should be for $P_{\rm s}$ to be determined mainly by 
%scattering within the atmosphere (this effect can be seen
%in Fig.~\ref{fig6}b, but it is not very obvious at this scale).

Figure~\ref{fig8} shows $F$ and $P_{\rm s}$ as functions of the
planet's orbital position angle for orbital inclination angles, 
$i$, ranging
from 0$^\circ$ (the orbit is seen face--on) to 90$^\circ$
(the orbit is seen edge--on). Given an inclination angle $i$,
an exoplanet can in principle be observed at phase angles,
$\alpha$, ranging from $90^\circ - i$ to $90^\circ + i$.
If the orbital position angle equals 0$^\circ$ or 360$^\circ$, 
$\alpha$ ranges from 
90$^\circ$ ($i=0^\circ$) to 0$^\circ$ ($i=90^\circ$).
If the orbital position angle equals 90$^\circ$ or 270$^\circ$, 
the planetary 
phase angle, $\alpha$, equals 90$^\circ$ (independent of $i$). 
If the orbital position angle equals 180$^\circ$, $\alpha$ ranges from 
90$^\circ$ ($i=0^\circ$) to 180$^\circ$ ($i=90^\circ$).
The curves in Fig.~\ref{fig8} clearly show that with increasing
orbital inclination angle, the variation of $F$ and $P_{\rm s}$
along the planetary orbit increases.
Interestingly, the orbital position angles where a planet is 
easiest to observe directly because it is furthest from its star 
(in angular distance) are those where $P_{\rm s}$ is largest
(namely, at orbital position angles equal to 90$^\circ$ and 270$^\circ$)
This emphasizes the strength of polarimetry for extrasolar
planet detection and characterization.
Incidentally, $P_{\rm s}$ is smallest for the inclination angles
and orbital position angles where it is the most difficult or even 
impossible to directly observe 
the planet, i.e. at large inclination angles and 
orbital position angles equal to 0$^\circ$, 180$^\circ$, or
360$^\circ$, when the planet
is close to, or even in front of or behind its star. 

%------------------------------------------------------------------
\subsection{Clear and cloudy planets with wavelength 
            dependent surface albedos}
\label{sect_wavelengthdependent}

%---------------------------------------------------------------------------
\subsubsection{Wavelength dependence}
\label{sect_wavelengthdependent1}

Figure~\ref{fig9} shows the wavelength dependence of the total 
flux, $F$, and degree of polarization, $P_{\rm s}$, of starlight that
is reflected by planets that are completely covered by, respectively,
ocean and deciduous forest, with atmospheres that are either clear, 
i.e.\ cloudless, or cloudy, i.e.\ that contain a homogeneous 
cloud layer. 
The cloud and the cloud particles have been described in 
Sect.~\ref{section_modelatmospheres}.
For comparison, we have also plotted 
$F$ and $P_{\rm s}$ of the clear white and black planets discussed
in Sect.~\ref{sect_wavelengthindependent}.
The planetary phase angle, $\alpha$, is 90$^\circ$.
First, we will discuss $F$ and $P_{\rm s}$ of the planets with
the clear atmospheres, and then those of the cloudy planets. 

The surface albedo of the clear planet that is covered by ocean
equals zero, regardless of wavelength 
(see Sect.~\ref{section_modelsurfaces}).
The differences between $F$ and $P_{\rm s}$ of the clear
black planet and $F$ and $P_{\rm s}$ of the clear, ocean 
covered planet (see Fig.~\ref{fig9}), are thus due to the 
specular reflecting interface between the model atmosphere 
and the ocean, which increases the total amount of light 
that is reflected back towards the observer.
In Fig.~\ref{fig9}a, the specular reflection increases $F$
in the continuum with about 10$\%$ at the short wavelengths
and with about 20$\%$ at the long wavelengths (where more
incoming starlight reaches the surface, because of the 
smaller atmospheric optical thickness). 
Although very difficult to see in Fig.~\ref{fig9}a, 
the influence of the specular reflection on $F$ is small 
within the gaseous absorption bands, because at those wavelengths,
little light reaches the surface and after a reflection
there, the top of the atmosphere again. 

As can be seen in Fig.~\ref{fig9}b, 
the specular reflection decreases $P_{\rm s}$ in the continuum 
with a few percentage points, 
because with the specular reflection, a fraction of the light
that is incident on the surface is reflected back towards
the atmosphere, adding mainly to the unpolarized flux
(at least at this phase angle). 
In gaseous absorption bands, the influence of the specular 
reflection on $P_{\rm s}$ is smaller than in the continuum, 
because at these wavelengths less light reaches the surface
and after a reflection there, the top of the atmosphere
again.

\begin{figure}
\vspace*{5.0cm}
\centering
\resizebox{8cm}{!}{\includegraphics{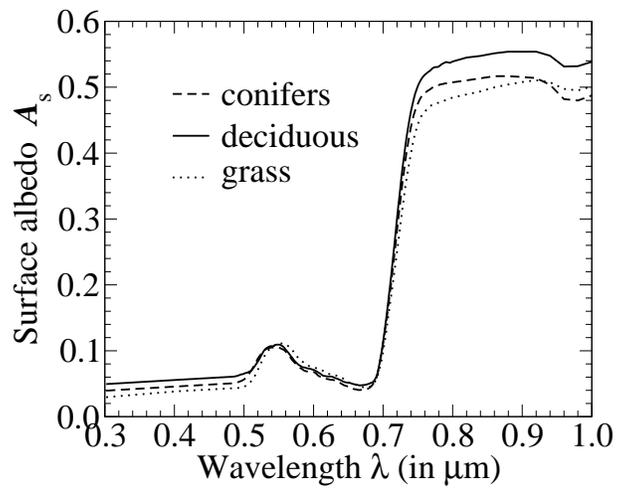}}
\vspace*{0.5cm}
\caption{The measured albedos of three types of common terrestrial 
         vegetation as functions of the wavelength: 
         conifers (dashed line), deciduous forest (solid line),
         and grass (dotted line) (data from the ASTER spectral 
         library). 
         For our model planets, we use 
         the albedo of deciduous forest to represent 
         vegetated surfaces.}
\label{fig4}
\end{figure}

\begin{figure}
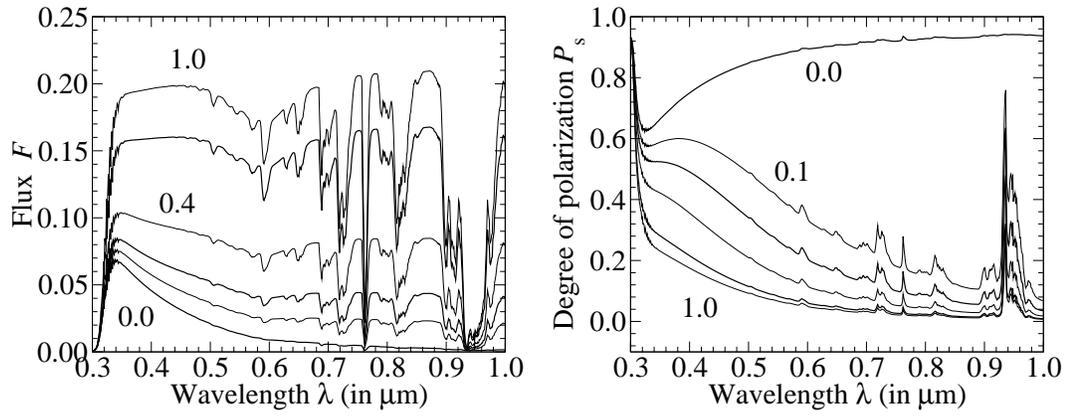

\vspace*{5.0cm}
\centering
\resizebox{14cm}{!}{\includegraphics{fig5a.eps} \hspace*{1.0cm}
                    \includegraphics{fig5b.eps} }
\vspace*{0.5cm}
\caption{The flux $F$ (left) and the degree of linear polarization
         $P_{\rm s}$ (right) of starlight reflected by model planets 
         with clear atmospheres
         and isotropically reflecting, completely depolarizing
         surfaces as functions of the wavelength,
         for various values of the (wavelength independent)
         surface albedo: 0.0, 0.1, 0.2, 0.4, 0.8, and 1.0.
         The planetary phase angle $\alpha$ is 90$^\circ$. 
         }
\label{fig5}
\end{figure}

\begin{figure}
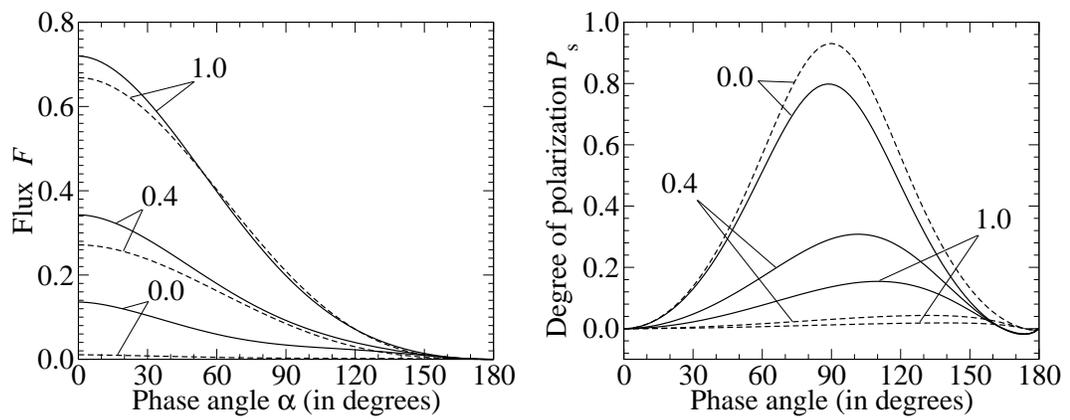

\vspace*{5.0cm}
\centering
\resizebox{14cm}{!}{\includegraphics{fig6a.eps} \hspace*{1.0cm}
                    \includegraphics{fig6b.eps} }
\vspace*{0.5cm}
\caption{The flux $F$ (left) and the degree of linear polarization
         $P_{\rm s}$ (right) of starlight reflected by the model 
         planets 
         with $A_{\rm s}=0.0$, 0.4, and 1.0 of Fig.~\ref{fig5}
         as functions of the phase angle $\alpha$.
         $F$ and $P_{\rm s}$ have been plotted for two 
         wavelengths: 0.44~$\mu$m
         (solid lines) and 0.87~$\mu$m (dashed lines).}
\label{fig6}
\end{figure}

\begin{figure}
\vspace*{5.0cm}
\centering
\resizebox{8cm}{!}{\includegraphics{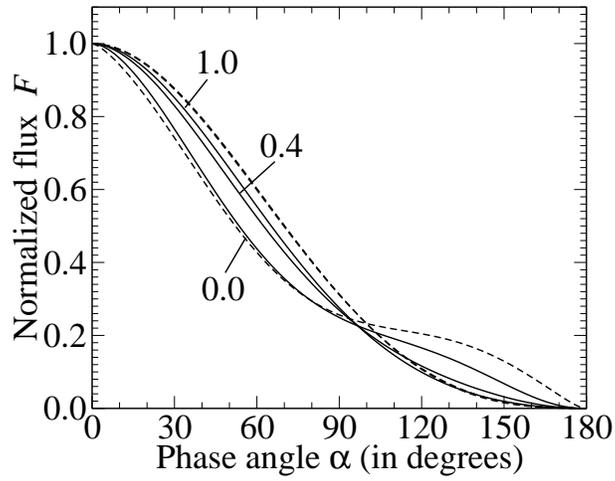}} 
\vspace*{0.5cm}
\caption{The fluxes of Fig.~\ref{fig6}a, normalized to 1.0 at
         $\alpha=0^\circ$. For \mbox{$\lambda=0.87~\mu$m} 
         (dashed lines),
         the curves for $A_{\rm s}=1.0$ and $A_{\rm s}=0.4$ are
         virtually indistinguishable, and follow the theoretical
         (normalized)
         curve expected for a Lambertian reflecting sphere
         \citep[see][]{1980vandeHulst,2006A&A...452..669S},
         i.e. $F(\alpha)=\frac{1}{\pi}\left(\sin\alpha+(\pi-\alpha) 
         \cos\alpha \right)$, with $\alpha$ in radians.}
\label{fig7}
\end{figure}

\begin{figure}
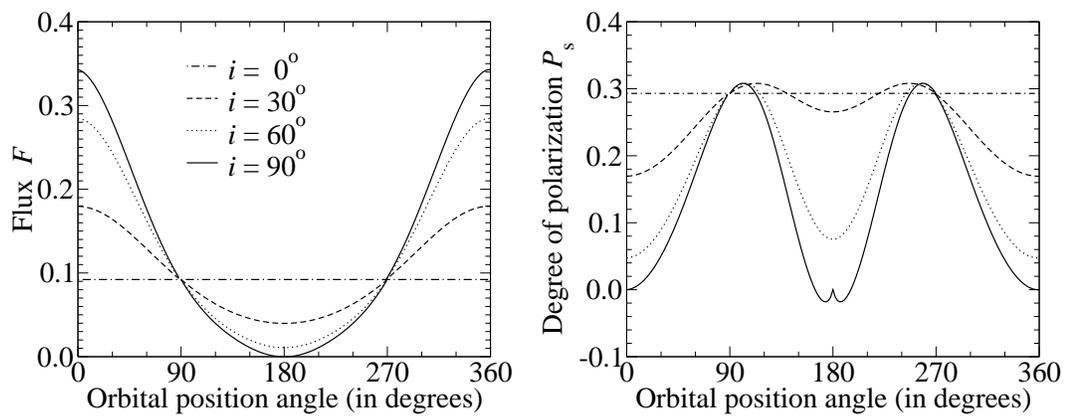

\vspace*{5.0cm}
\centering
\resizebox{14cm}{!}{\includegraphics{fig8a.eps} \hspace*{1.0cm}
                    \includegraphics{fig8b.eps} }
\vspace*{0.5cm}
\caption{$F$ (left) and $P_{\rm s}$ (right) of starlight with
         $\lambda=0.44~\mu$m that is reflected by 
         the model planet with $A_{\rm s}=0.4$
         as functions of the orbital position angle for the 
         following orbital
         inclination angles $i$: 0$^\circ$ (dot-dashed lines), 
         30$^\circ$ (dashed lines), 60$^\circ$ (dotted lines), 
         and 90$^\circ$ (solid lines).}
\label{fig8}
\end{figure}

\begin{figure}
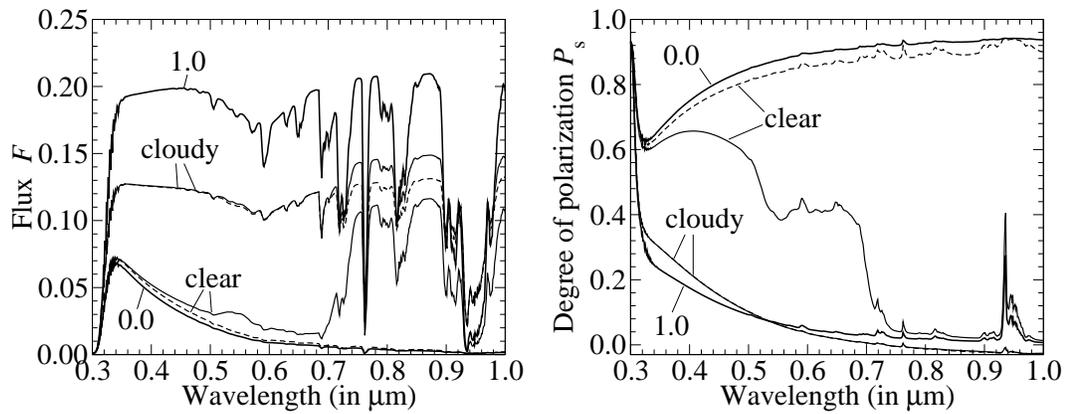

\vspace*{5.0cm}
\centering
\resizebox{14cm}{!}{\includegraphics{fig9a.eps} \hspace*{1.0cm}
                    \includegraphics{fig9b.eps} }
\vspace*{0.5cm}
\caption{The wavelength dependent flux, $F$, (left) and 
         degree of polarization, $P_{\rm s}$, (right) of starlight
         that is reflected by clear and cloudy horizontally
         homogeneous model planets with surfaces covered by
         deciduous forest (thin solid lines),
         a specular reflecting ocean (thin dashed lines).
         Note that the lines pertaining to $P_{\rm s}$ of
         the cloudy atmospheres are virtually indistinguishable
         from each other.
         For comparison, we have also included the spectra
         of the clear model planets with surface albedos
         equal to 0.0 and 1.0 (thick solid lines), shown 
         before in Fig.~\ref{fig5}.
         The planetary phase angle, $\alpha$, is 90$^\circ$. 
         }
\label{fig9}
\end{figure}

The wavelength variation of the continuum $F$ and $P_{\rm s}$
pertaining to the clear, forest--covered planet
reflects the wavelength variation 
of the surface albedo (see Fig.~\ref{fig4}) except at
the shortest wavelengths. There, $F$ and $P_{\rm s}$ are mainly 
determined by the light that has been scattered by the gaseous 
molecules within the planet's atmosphere, because (1) the 
atmospheric scattering optical thickness increases with decreasing
wavelength, and (2) the surface albedo is only about 0.05
at the shortest wavelengths (see Fig.~\ref{fig4}).
At longer wavelengths, the characteristic reflection by 
chlorophyll (around 0.54~$\mu$m) 
and the red edge (longwards of 0.7~$\mu$m) 
can easily be recognized both in $F$ and in $P_{\rm s}$.
The red edge in flux spectra of the Earth
has been detected by instruments onboard the Galileo mission
on its way to Jupiter \citep[][]{1993Natur.365..715S}, 
and from the ground it has been observed 
and in some cases modeled by different research groups
\citep[see e.g.][]{2006A&A...460..617H,2006ApJ...651..544M,
2006AsBio...6..881T,2006AsBio...6...34T,
2005AsBio...5..372S,2002ApJ...574..430W,
2002A&A...392..231A}
in spectra of Earthshine,
the sunlight that has first
been reflected by the Earth and then by the moon, and that
can be observed on the moon's nightside.
Interestingly, the reflection by chlorophyll leaves a much
stronger signature in $P_{\rm s}$ than in $F$, because in 
this wavelength
region $P_{\rm s}$ appears to be very sensitive to small changes
in $A_{\rm s}$, as can also be seen in Fig.~\ref{fig5}b.

Adding a cloud layer to the atmosphere of a planet covered with either
vegetation or ocean, increases $F$ across the whole wavelength
interval (see Fig.~\ref{fig9}a).
A discussion on the effects of different types of clouds on 
flux spectra of light reflected by exoplanets is given by 
\citet{2006AsBio...6..881T,2006AsBio...6...34T}.
Our simulations show that although the cloud layers of the 
two cloudy planets have a
large optical thickness 
(i.e. 10 at $\lambda=0.55$~$\mu$m, 
as described in Sect.~\ref{section_modelatmospheres})
both cloudy planets in Fig.~\ref{fig9}a are darker than 
the white planet with the clear atmosphere (the flux of
which is also plotted in Fig.~\ref{fig9}a). 
The cloud particles themselves are only slightly absorbing 
(see Sect.~\ref{section_modelatmospheres}).
Apparently, on the cloudy planets, a
significant amount of incoming starlight is diffusely
transmitted through the cloud layer (through multiple
scattering of light) and then absorbed
by the planetary surface (the albedos of the ocean and the
forest are smaller than 1.0). 
Thus, even with an optically thick cloud,
the albedo of the planetary surface still influences
the light that is reflected by the planets, and approximating
clouds by isotropically or anisotropically
reflecting surfaces, without regard for what is underneath,
as is sometimes done 
\citep[see e.g.][]{2006ApJ...651..544M,2002ApJ...574..430W}
is not appropriate.
Assuming a dark surface beneath
scattering clouds with non-negligible optical thickness
\citep[][]{2006AsBio...6..881T,2006AsBio...6...34T}
will lead to too dark planets.
The influence of the surface albedo is 
in particularly clear for the cloudy planet that is covered with 
vegetation: longwards of 0.7~$\mu$m, the continuum
flux of this planet still shows the vegetation's red edge.
The visibility of the red edge through optically thick clouds
strengthens the detectability of surface biosignatures in the
visible wavelength range as discussed by 
\citep[][]{2006AsBio...6..881T}, whose numerical simulations 
showed that, averaged over the daily time scale, Earth's land 
vegetation would be visible in disk-averaged
spectra, even with cloud cover, and even without accounting for
the red edge below the optically thick clouds. 
Note that the vegetation's albedo signature due to chlorofyll,
around 0.54~$\mu$m, also shows up in Fig.~\ref{fig9}a, 
but hardly distinguishable.

The degree of polarization, $P_{\rm s}$, of the cloudy planets
is low compared to that of planets with clear atmospheres, except at
short wavelengths.
The reasons for the low degree of polarization of the
cloudy planets are
(1)~the cloud particles strongly increase the amount of
multiple scattering of light within the atmosphere,
which decreases the degree of polarization,
(2)~the degree of polarization of light that is 
{\em singly} scattered by the cloud particles is generally lower than 
that of light singly scattered by gaseous molecules, especially
at single scattering angles around 90$^\circ$ (see Fig.~\ref{fig3}b),
and
(3)~the direction of polarization of light singly scattered
by the cloud particles is opposite to that of light singly scattered
by gaseous molecules (see Fig.~\ref{fig3}b).
Thanks to the latter fact, the continuum $P_{\rm s}$ 
of the cloudy planets is negative (i.e. the direction of 
polarization is perpendicular to the terminator) 
at the longest wavelengths 
(about -0.03 or 3~$\%$ for $\lambda>0.73~\mu$m in Fig.~\ref{fig9}b). 
At these wavelengths, the atmospheric molecular scattering optical 
thickness is negligible compared to the optical thickness 
of the cloud layer, and therefore almost all
of the reflected light has been scattered by cloud particles.

Unlike in the flux spectra, the albedo of the surface below 
a cloudy atmosphere leaves almost no trace in $P_{\rm s}$
of the reflected light:
the red edge of the vegetation hardly leads to a difference between
$P_{\rm s}$ of the cloudy planets (Fig.~\ref{fig9}b). 
In particular, at 1.0~$\mu$m, $P_{\rm s}$ of the cloudy,
vegetation--covered planet is -0.030 (-3.0$\%$), while $P_{\rm s}$ of the
cloudy, ocean--covered planet is -0.026 (-2.6$\%$). 
The reason of the insensitivity of $P_{\rm s}$ of these two 
cloudy planets to the surface albedo is that  
the light reflected by the surfaces in our models
adds mainly unpolarized light to the atmosphere, in a wavelength
region where $P_{\rm s}$ is already very low because of the clouds. 

The cloud layer has interesting effects on the strengths of 
the absorption bands of O$_2$ and H$_2$O both in $F$ and in
$P_{\rm s}$: because the cloud
particles scatter light very efficiently, their presence
strongly influences the average pathlength of a photon within the
planetary atmosphere. 
At wavelengths where light is absorbed by atmospheric gases,
clouds thus strongly change the fraction of light that
is absorbed, and with that the strength of the absorption band.
These are well-known effects in Earth remote-sensing;
in particular the O$_2$ $A$-band is used 
to derive e.g. cloud top altitudes and/or
cloud coverage within a ground pixel 
\citep[see e.g.][]{1994JGR....9914481K,
1991JApMe..30.1245F,
1991JApMe..30.1260F,1967JAtS...24...63S,
2000JGR...10522379S}, because oxygen is well-mixed within
the Earth's atmosphere.
In general, clouds will decrease the relative depth (i.e. with
respect to the continuum) of absorption bands in
reflected flux spectra (see Fig.~\ref{fig9}a), because they
shield the absorbing gases that are below them. However,
because of the multiple scattering within the clouds, 
the absorption bands will be deeper than expected when using
a reflecting surface to mimic the clouds.
For example, the discrepancy between absorption band depths 
in Earth-shine flux observations and model simulations as shown by 
\citet{2006ApJ...651..544M}, with the observation yielding
e.g. a deeper O$_2$-A band than the model can fit, 
can be due to neglecting (multiple)
scattering within the clouds, as \citet{2006ApJ...651..544M} 
themselves also remark.

Another source for differences between absorption
band depths in observed and modeled flux spectra could be that
when modeling albedo and/or flux spectra, the state of 
polarization of the light is usually neglected.
\citet[][]{2005A&A...444..275S} showed for Jupiter-like 
extrasolar planets that neglecting polarization can 
lead to errors of up to 10~$\%$
in calculated geometric albedos, and that in particular the depths
of absorption bands are affected, because the error in the continuum
is usually much larger than the error in the deepest part of 
the absorption band. 
In Sect.~\ref{section_summary}, we will show that 
neglecting polarization does not significantly change the
depth of the absorption bands in the flux spectra of these
Earth-like model planets.

In the polarization spectra (Fig.~\ref{fig9}b), the effects of 
clouds on the strength of the absorption band features 
are more complicated than in flux spectra, because 
the absorption not only changes the amount of multiple
scattering that takes place in the atmosphere, but also
the altitude where the reflected
light has obtained its state of polarization; in an
inhomogeneous atmosphere, like a cloudy one, different particles
at different altitudes can leave the light they scatter in
different states of polarization (this also affects the 
scattered fluxes, but less so) \citep[see e.g.][]{1999JGR...10416843S}.
In Fig.~\ref{fig9}b, $P_{\rm s}$ slightly increases within
absorption bands in wavelength regions where the continuum
$P_{\rm s}$ is positive ($\lambda<0.73~\mu$m), 
whereas $P_{\rm s}$ slightly 
decreases (in absolute sense) within absorption bands
in wavelength regions where the continuum $P_{\rm s}$ 
is negative ($\lambda>0.73~\mu$m). 
In these cloudy model atmospheres,
both the increase and the decrease (in absolute sense)
of $P_{\rm s}$ in absorption bands are due to (1)~a decrease of
multiple scattering (which also takes place in purely gaseous
model atmospheres), and (2)~an increase of the relative amount
of photons that are scattered by gaseous molecules instead
of by cloud particles, since the latter are located in the lower
atmospheric layers.

The change of the strength of an absorption band in $F$ or 
$P_{\rm s}$ due to the presence of a cloud layer,
depends strongly on the altitude of the cloud layer,
its optical thickness, the cloud coverage , the mixing ratio and
the vertical distribution of the absorbing gas.
For example, in both the flux and the polarization spectra of 
the cloudy planets in Fig.~\ref{fig9}, the absorption bands of H$_2$O 
are weak compared to the same bands in the spectra of the 
cloudless planets, because most of the H$_2$O
is located below the clouds.
The absorption bands of O$_2$ are also weaker for the 
cloudy planets than for the cloudless planets, although the
influence of the clouds on these absorption bands is less 
strong than on the bands of H$_2$O, simply because O$_2$
is well--mixed throughout the atmosphere, and thus not primarily
located below the clouds, like H$_2$O.

\begin{figure}
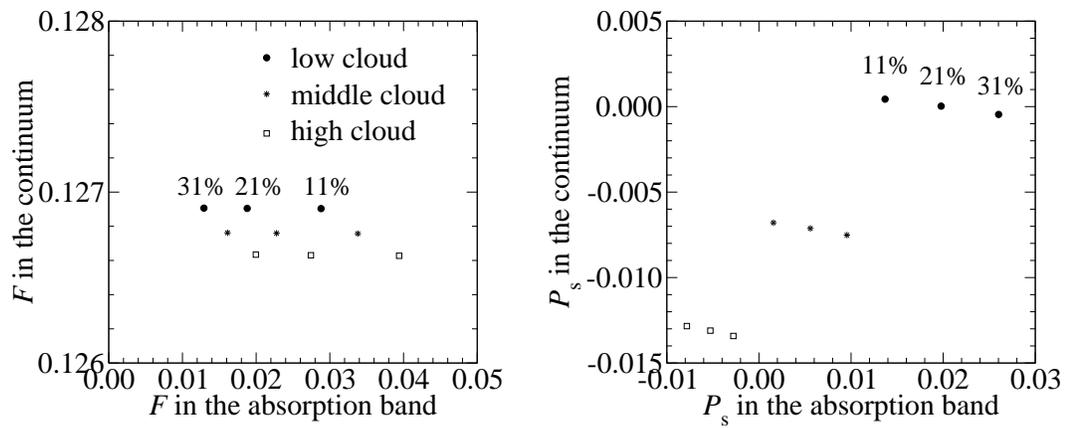

\vspace*{5.0cm}
\centering
\resizebox{14cm}{!}{\includegraphics{fig10a.eps} \hspace*{2.5cm}
                    \includegraphics{fig10b.eps} }
\vspace*{0.5cm}
\caption{The flux $F$ (left) and the degree of linear polarization
         $P_{\rm s}$ (right) of starlight reflected by
         cloudy ocean planets, at $\lambda= 0.762~\mu$m, 
         {\em with} absorption by O$_2$ (along the $x$-axes) and
         {\em without} absorption by O$_2$ (along the $y$-axes).   
         The top of the cloud layer was located at 
         802~hPa (the 'low cloud'; the nominal altitude), 628~hPa
         (the 'middle cloud'), or 487~hPa (the 'high cloud').
         The O$_2$ mixing ratio was 11$\%$, 21$\%$ 
         (the nominal value), or 31$\%$.
         The planetary phase angle $\alpha$ is 90$^\circ$. 
         }
\label{fig10}
\end{figure}

When a terrestrial type extrasolar planet will be discovered,
it will of course be extremely interesting to try to identify
oxygen in the planet's atmosphere, and in particular 
to determine the oxygen mixing ratio
from absorption bands such as the O$_2$ $A$-band.
As discussed above, the depth of such an absorption 
band will depend not only on the absorber's mixing ratio,
but also on the cloud cover.
As an example,
Fig.~\ref{fig10} shows the influence of the cloud
top altitude and the O$_2$ mixing ratio on the depth of the
O$_2$-$A$ absorption band, both for the flux and 
the degree of polarization of the reflected starlight,
for $\alpha=90^\circ$.
The figure shows the results of numerical calculations
for model planets covered by ocean and with cloud 
layers of optical
thickness 10 (at 0.55~$\mu$m) placed with their tops
at, respectively, 802~hPa, 628~hPa, and 487~hPa,
and with oxygen mixing ratios of, respectively,
11~$\%$, 21~$\%$, and 31~$\%$.  
From Fig.~\ref{fig10}a, it is clear that the continuum
flux is independent of the O$_2$ mixing ratio,
as it should be, and that it is virtually
independent of the cloud top altitude (note the 
vertical scale). The latter is easily understood 
by realizing that at these wavelengths, the gaseous
molecular scattering optical thickness above the
different cloud layers is negligible compared to 
the scattering optical thickness of the cloud layers
themselves.
Not surprisingly, the flux {\em in} the absorption 
band increases significantly with a decreasing O$_2$ 
mixing ratio and with increasing cloud
top altitude (i.e. decreasing cloud top pressure).
Concluding, the depth of the O$_2$ $A$--band in planetary 
flux spectra depends on the O$_2$ mixing ratio as well as 
on the cloud top altitude, and it can thus not be used
for deriving the O$_2$ mixing ratio if the cloud
top altitude is unknown or vice versa.
Figure~\ref{fig10}b shows the relation between $P_{\rm s}$
in the continuum and in the absorption band, 
and the cloud top altitude and oxygen mixing ratio. 
Apparently, the continuum $P_{\rm s}$ depends significantly
on the cloud top altitude, because the degree of polarization
is very sensitive to even small amounts of gaseous 
molecules, and depends only very slightly on the oxygen mixing ratio.
The degree of polarization {\em in} the absorption band 
depends on the oxygen mixing ratio as well as on the
cloud top altitude.
Concluding, the strength of the O$_2$ $A$ band in planetary
polarization spectra can be used to derive both the cloud
top altitude and the oxygen mixing ratio.
The influence of e.g. broken cloud layers, and clouds at
different altitudes on $F$ and $P_{\rm s}$, will be subject
for later studies.

\subsubsection{Phase angle dependence}

\begin{figure}
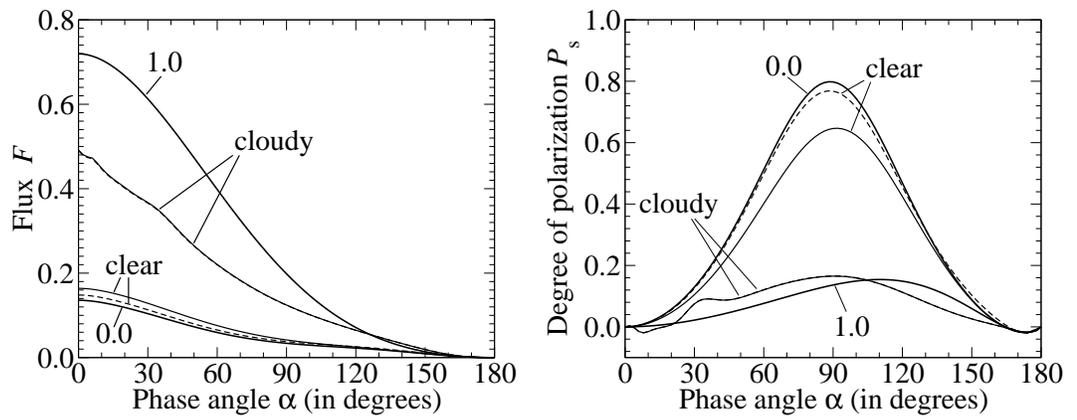

\vspace*{5.0cm}
\centering
\resizebox{14cm}{!}{\includegraphics{fig11a.eps} \hspace*{1.0cm}
                    \includegraphics{fig11b.eps}}
\vspace*{0.5cm}
\caption{The flux $F$ (left) and the degree of linear polarization
         $P_{\rm s}$ (right) of starlight reflected by the model 
         planets of Fig.~\ref{fig9}
         as functions of the phase angle $\alpha$, for 
         $\lambda=0.44$~$\mu$m. The thin solid line pertains to
         the vegetation--covered planet, and the thin dashed line 
         to the ocean--covered planet. The lines pertaining to
         the cloudy planets are virtually indistinguishable
         from each other.}
\label{fig11}
\end{figure}

\begin{figure}
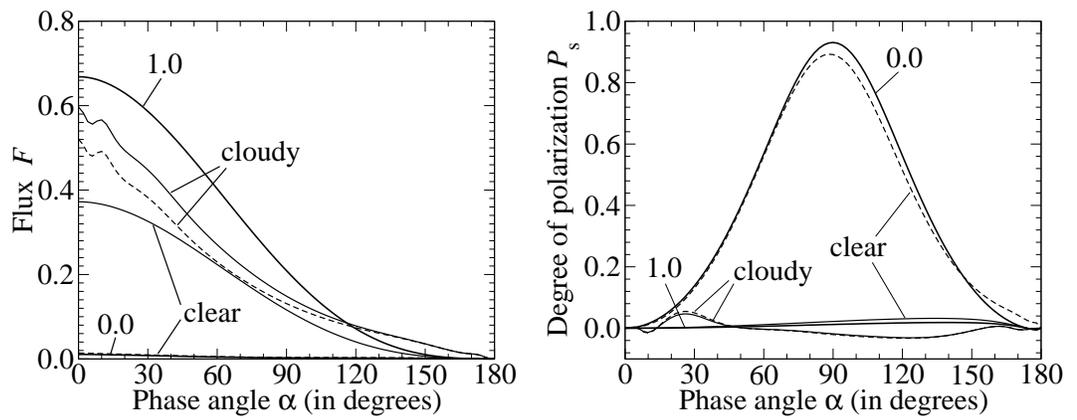

\vspace*{5.0cm}
\centering
\resizebox{14cm}{!}{\includegraphics{fig12a.eps} \hspace*{1.0cm}
                    \includegraphics{fig12b.eps}}
\vspace*{0.5cm}
\caption{Same as Fig.~\ref{fig11}, except for $\lambda=0.87$~$\mu$m.
         The lines for $P_{\rm s}$ pertaining to the cloudy planets 
	 are virtually indistinguishable from each other, except 
	 at phase angles between 20$^\circ$ and 40$^\circ$.} 
\label{fig12}
\end{figure}

Figures~\ref{fig11} and~\ref{fig12} show the phase angle 
dependence of $F$ and $P_{\rm s}$ of the starlight that is 
reflected by the clear and cloudy, ocean and vegetation--covered 
planets appearing in the previous section 
(see Sect.~\ref{sect_wavelengthdependent1} and
Fig.~\ref{fig9}) at $\lambda=0.44~\mu$m (Fig.~\ref{fig11}) 
and at $\lambda=0.87~\mu$m (Fig.~\ref{fig12}), respectively. 
Remember that at phase angle $\alpha=0^\circ$, the fluxes
plotted in Figs.~\ref{fig11}a and~\ref{fig12}a are just the
planets' geometric albedos, $A_{\rm G}$, at those wavelengths.
For the clear, ocean--covered planet, $A_{\rm G}$ is 0.15 at
$\lambda=0.44~\mu$m, and 0.014 at $0.87~\mu$m.
For the clear, vegetation--covered planet, $A_{\rm G}$ is 0.16 at
$0.44~\mu$m, and 0.37 at $0.87~\mu$m.
For the cloudy, ocean--covered planet, $A_{\rm G}$ is 0.49 at
$0.44~\mu$m, and 0.52 at $0.87~\mu$m,
and for the cloudy, vegetation--covered planet, $A_{\rm G}$ 
is 0.49 at $0.44~\mu$m (like for the cloudy, ocean--covered
planet), and 0.60 at $0.87~\mu$m.

For the two cloudy planets, 
the reflected flux is not a smoothly decreasing function of
phase angle, but instead shows some bumpy features
near $\alpha=5^\circ$ and 30$^\circ$ for $\lambda=0.44~\mu$m
(Fig.~\ref{fig11}a), and near 10$^\circ$ and 35$^\circ$ 
for $\lambda=0.87~\mu$m (Fig.~\ref{fig12}a).
These angular features trace back to features 
in the single scattering phase function of the cloud particles
(at scattering angles, $\Theta$, larger than 140$^\circ$)
(see Fig.~\ref{fig3}a).
Because at $\lambda=0.44~\mu$m, the gaseous molecules above
the cloud layer scatter light more efficiently than at
$\lambda=0.87~\mu$m, the angular features in $F$ are
more subdued at 0.44~$\mu$m than at 0.87~$\mu$m.
Interestingly, at phase angles near 60$^\circ$ and for
$\lambda=0.87~\mu$m (Fig.~\ref{fig12}a), the
cloudy, ocean--covered planet is about as bright as the clear,
vegetation--covered planet.
Apart from the high surface albedo of the vegetation at this
wavelength (Fig.~\ref{fig4}),
this can partly be attributed to the single scattering
phase function of the cloud particles, too,
because this function has a broad minimum between 
$\Theta=90^\circ$ and $\Theta=120^\circ$ (see Fig.~\ref{fig3}a).
Note further that only at the largest phase angles, 
i.e. $\alpha > 120^\circ$,
the cloudy planets are brighter than the clear planet 
with the surface albedo equal to 1.0. This is in particularly
obvious at $\lambda=0.87~\mu$m (Fig.~\ref{fig12}a),
and is due to the strong forward scattering peak in the
single scattering phase functions of the cloud particles
(Fig.~\ref{fig3}a). 
At these large phase angles, the light that is reflected 
by the cloudy planets has not reached the surface, and the
reflected flux is therefore independent of the surface albedo 
(see Fig.~\ref{fig12}a).

The phase angle dependence of the degree of linear polarization,
$P_{\rm s}$, of the light reflected by the two cloudy planets 
(Figs.~\ref{fig11}b and \ref{fig12}b) shows, for $\alpha < 40^\circ$, 
angular features 
that are due to angular features in the single scattering 
polarization phase function of the cloud particles 
(see Fig.~\ref{fig3}b), just like the phase angle
dependence of the reflected flux
(Figs.~\ref{fig11}a and \ref{fig12}a). 
At $\lambda=0.44~\mu$m, $P_{\rm s}$ is determined not only
by light scattered by the cloud particles, but also by 
light scattered by the gaseous molecules, whereas at
$\lambda=0.87~\mu$m, $P_{\rm s}$ is predominantly determined
by the cloud particles. As a result, at $\lambda=0.87~\mu$m,
the characteristic polarization signatures of light scattered 
by the cloud particles (i.e. the angular features at 
$\alpha < 40^\circ$ and also the negative values for 
60$^\circ$~$<$~$\alpha$~$<$~160$^\circ$) are much stronger
than at $\lambda=0.44~\mu$m (taking into account the 
wavelength dependence of the single scattering features
themselves, cf. Fig.~\ref{fig3}).

The angular features around $\alpha=32^\circ$
for $\lambda=0.44~\mu$m, and around $\alpha=25^\circ$ for
$\lambda=0.87~\mu$m, pertain to the so-called primary rainbow
feature. At $\lambda=0.44~\mu$m, the degree of polarization 
of this feature is about 0.1 (10$\%$) and at $\lambda=0.87~\mu$m, about 
0.06 (6$\%$). This is much lower than the 20$\%$ predicted, for
a completely cloudy Earth, by \citet{2007AsBio...7..320B}.
This discrepancy is probably due to the fact that 
\citet{2007AsBio...7..320B} arrives at his disk-integrated 
value by integrating local 
observations by the POLDER-satellite instrument 
\citep[][]{1994Deschamps}, while our multiple scattering
calculations and disk-integration method 
take into account the variation of illumination
and viewing angles across the planetary disk. In addition,
differences in single-scattering properties of cloud
particles (which depend on the particle size distribution
and composition) and the optical thickness of the clouds
will influence the strength of the feature,
although probably not more than a few percent.

At neither of the two wavelengths, $P_{\rm s}$ of the cloudy 
planets is sensitive to the albedo of the surface below the
clouds (Figs.~\ref{fig11}b and \ref{fig12}b); 
only at 0.87~$\mu$m and for $\alpha~<~40^\circ$,  
$P_{\rm s}$ of the cloudy, ocean--covered planet is at most 
0.01 (1$\%$) larger than $P_{\rm s}$ of the cloudy, vegetation--covered
planet, mainly because of the darkness of the ocean. 

For the clear planets with the Lambertian reflecting surfaces, 
the phase angle dependence of $P_{\rm s}$
at both wavelengths (Figs.~\ref{fig11}b and~\ref{fig12}b) 
is similar to that shown in Fig.~\ref{fig6}b, 
taking into account the differences in surface albedo. 
Compared with the clear, black planet,
the specular reflecting surface of the clear, ocean--covered 
planet decreases $P_{\rm s}$ at all phase angles (with
at most 0.04 at $\alpha=90^\circ$ and $\lambda=0.87~\mu$m), 
except at phase angles larger than about 130$^\circ$ at 
$0.44~\mu$m, and about 145$^\circ$ at $0.87~\mu$m. 

%------------------------------------------------------------------
\subsection{Clear and cloudy planets with horizontal 
inhomogeneities} 
\label{sect_mixtures}

\subsubsection{Wavelength dependence}

In Sects.~\ref{sect_wavelengthindependent} 
and~\ref{sect_wavelengthdependent}, we have presented numerically 
simulated fluxes and degrees of polarization of light reflected by 
{\em horizontally homogeneous} planets. In this section, we will 
show results for {\em quasi horizontally inhomogeneous} planets, 
using Eq.~\ref{eq_weightedsum} and the flux vectors calculated
for the clear and cloudy, ocean-- and vegetation--covered
planets presented in Sect.\ref{sect_wavelengthdependent}
and Figs.~\ref{fig9}, \ref{fig11}, and~\ref{fig12}.
The resulting spectra can be thought of as 
the Earth observed as if it were an exoplanet, 
and using an integration time of at least a day.
Since we use a weighted sum of homogeneous planets,
our spectra might differ from those obtained using a model
planet covered by continents and oceans, even if the 
coverage fractions are the same. Our spectra
will, however, give a good estimate of what might be expected. 

Although of course endless combinations of these flux vectors
can be made, we will limit ourselves here to Earth--like
combinations, and leave other combinations and retrieval 
algorithms
for subsequent studies. In addition, including more types of 
surface coverage, different types of clouds, and e.g. 
different cloud coverages
for different surface types, would add details to 
the modelling results that are beyond the scope of this article.
See e.g. \citet{2006AsBio...6...34T,2006AsBio...6..881T}
for examples of flux spectra for different cloud types and
surface coverages.

\begin{figure}
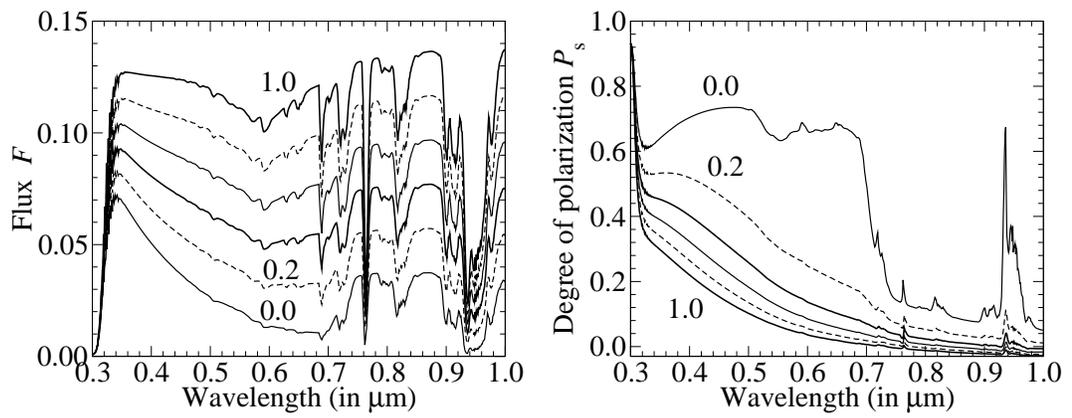

\vspace*{5.0cm}
\centering
\resizebox{14cm}{!}{\includegraphics{fig13a.eps} \hspace*{1.0cm}
                    \includegraphics{fig13b.eps} }
\vspace*{0.5cm}
\caption{Similar to Fig.~\ref{fig9}, except for quasi horizontally
         inhomogeneous planets with 70$\%$ of their surfaces
         covered by specular reflecting ocean and 30$\%$ by
         deciduous forest. The cloud coverage ranges from 0.0
         to 1.0, in steps of 0.2.
         The planetary phase angle, $\alpha$, is 90$^\circ$. 
         }
\label{fig13}
\end{figure}

Figure~\ref{fig13} shows the flux and degree of polarization
of light reflected by an exoplanet that has, like the Earth,
70~$\%$ of its surface covered by a specular reflecting ocean
and 30~$\%$ by deciduous forest. The cloud coverage ranges
from 0.0 (a clear atmosphere) to 1.0 (a completely overcast
sky) in steps of 0.2. 
Recall that the mean global cloud coverage of
the Earth is about 0.67 \citep{Rossow1996}.
Note that to simulate $F$ and $P_{\rm s}$ 
of the cloudy fractions of the planets, we combined
the cloudy, ocean--covered planets with the cloudy,
forest--covered planets. In Eq.~\ref{eq_weightedsum},
we thus used $N=4$.

The main, not surprising difference between the flux of the 
clear, quasi horizontally inhomogeneous planet
from Fig.~\ref{fig13}a, and that of the clear, horizontally homogeneous
forest--covered planet in Fig.~\ref{fig9}a, is that the red edge 
(for $\lambda > 0.7~\mu$m) that is
characteristic for reflection by vegetation, is much less 
strong when 70~$\%$ of the
planet is covered by ocean than when the whole planet is
covered by vegetation (the red edge continuum is approximately
0.035 for the inhomogeneous planet and 0.11 for the homogeneous one).
The (black) ocean all but removes the local maximum in $F$ 
at green wavelengths (between 0.5 and 0.6~$\mu$m) 
which is due to chlorophyll in the vegetation.
In $P_{\rm s}$ (Fig.~\ref{fig13}b), the ocean somewhat 
changes the spectral shape
of the red edge feature, and decreases its depth by about 
0.04 (at $\lambda=0.87~\mu$m) when compared to Fig.~\ref{fig9}b.
The ocean signficantly changes the spectral feature in $P_{\rm s}$
that is due to chlorofyll: for the clear, completely forest--covered
planet (Fig.~\ref{fig9}b), the minimum $P_{\rm s}$ across this
feature is 0.38 while it is 0.63 for the clear planet in 
Fig.~\ref{fig13}b.
Finally, in $P_{\rm s}$, the gaseous absorption bands are stronger 
for the clear planet covered by ocean and forest, than for the
clear, forest--covered planet.

From Fig.~\ref{fig13}a, it is clear that $F$ in the continuum is 
very sensitive to the cloud coverage. The continuum $P_{\rm s}$
is also sensitive to the cloud coverage, but this sensitivity
decreases with increasing cloud coverage, in particularly at
the longer wavelengths.  

\begin{figure}
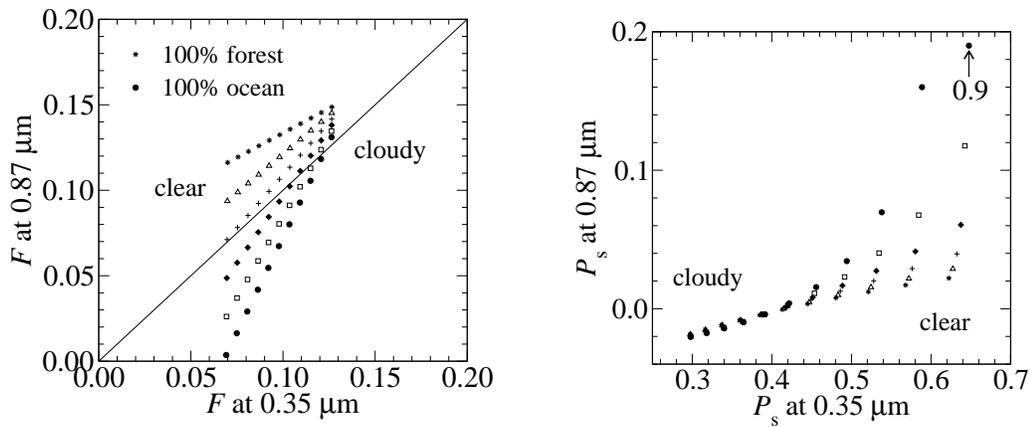

\vspace*{5.0cm}
\centering
\resizebox{14cm}{!}{\includegraphics{fig14a.eps} \hspace*{2.5cm}
                    \includegraphics{fig14b.eps} }
\vspace*{0.5cm}
\caption{The relations between $F$ (left) and $P_{\rm s}$ (right)
         in the blue ($\lambda= 0.35~\mu$m) and the near--infrared
         ($\lambda=0.87~\mu$m) for planets with ratios between
         ocean and forest--surface coverage ranging from 0.0
         (100~$\%$ covered by forest, indicated by the stars),
         to 0.0 (100~$\%$ covered by ocean, indicated by the 
         black circles), with steps of 0.2.  
         In addition, the cloud coverage ranges from 0.0
         to 1.0, in steps of 0.1.
         The planetary phase angle, $\alpha$, is 90$^\circ$. 
         Note that we have included the line 
         $F$(0.87~$\mu$m)=$F$(0.35~$\mu$m) in the graph on the
         left, and  
         that at $\lambda=0.87~\mu$m, $P_{\rm s}$ of the 
         clear, 100~$\%$ ocean planet falls of the vertical 
         figure scale; its value is 0.9. 
         }
\label{fig14}
\end{figure}

To get more insight into the sensitivity of the shape of 
the flux and polarization spectra
to the surface and the cloud coverage, we have plotted in 
Fig.~\ref{fig14}, $F$ and $P_{\rm s}$ in the near--infrared
continuum ($\lambda=0.87~\mu$m) against $F$ and $P_{\rm s}$
in the blue continuum ($\lambda=0.35~\mu$m) for planets
with surface coverage ratios ranging from 0.0 (100~$\%$ forest) to 
1.0 (100~$\%$ ocean), in steps of 0.2.
The cloud coverage ranges from 0.0 (a clear planet) to 1.0
(a completely cloudy planet).
Looking at the reflected fluxes (Fig.~\ref{fig14}a),
it can be seen that for a given cloud coverage, 
$F$ in the blue is virtually independent of the 
surface coverage (ocean or forest), while in the near--infrared, 
the sensitivity of $F$ to the surface coverage depends
strongly on the cloud coverage: it is relatively
large when the cloud coverage is small, and
relatively small when the cloud coverage is large, as could 
be expected.
Fig.~\ref{fig14}a shows that our completely
cloudy planets are somewhat brighter in the near--infrared
than in the blue (assuming a wavelength independent stellar
flux, or, after correcting observations for the incoming
stellar flux).
Looking at the degree of polarization of the reflected
fluxes (Fig.~\ref{fig14}b), it is clear that the larger the cloud
coverage, the smaller the dependence of $P_{\rm s}$ 
at both 0.35~$\mu$m and 0.87~$\mu$m on the surface coverage.
For a given cloud coverage smaller than about 0.5, 
$P_{\rm s}$ can be seen to depend on the surface coverage.
In particular, the larger the fraction of ocean on the
planet, the larger $P_{\rm s}$ in the near--infrared.
In the blue, $P_{\rm s}$ also increases with increasing
fraction of ocean coverage, but only slightly so.

It is important to remember that the precise
location of the data points in Fig.~\ref{fig14}, and thus the
retrieval opportunities, will
depend on the physical parameters of the model cloud,
such as the cloud optical thickness, the microphysical
properties of the cloud particles, and the altitude
of the cloud. We will explore such dependencies in 
later studies.

\subsubsection{Phase angle dependence and diurnal variation}

As a horizontally inhomogeneous planet like the Earth rotates
around its axis, the surface fraction of, for example, ocean that
is turned towards a distant observer will vary during the day
(except when the observer is located precisely above one of the
planet's geographic poles). Assuming a planet with a surface 
that is covered only by land and water, the diurnal variation of 
the distribution of 
land and water across the part of the planetary disk that 
is illuminated and turned towards a distant observer
depends on many factors, such as the actual distribution
of land and water across the planet,
the sub-observer latitude, the planet's phase angle, and 
the location of the terminator (the division between day
and night on the planet, that will depend on the obliquity
of the planet and the time of year).
The variation that can actually be observed, would, of course,
also depend on the cloud cover.
Calculated diurnal variations of fluxes of Earth-like exoplanets 
have been presented by \citet{2001Natur.412..885F}, using a 
single scattering Monte Carlo radiative transfer code and 
horizontally inhomogeneous model planets, and by
\citet{2006AsBio...6..881T}.

\begin{figure}
\vspace*{5.0cm}
\centering
\resizebox{8cm}{!}{\includegraphics{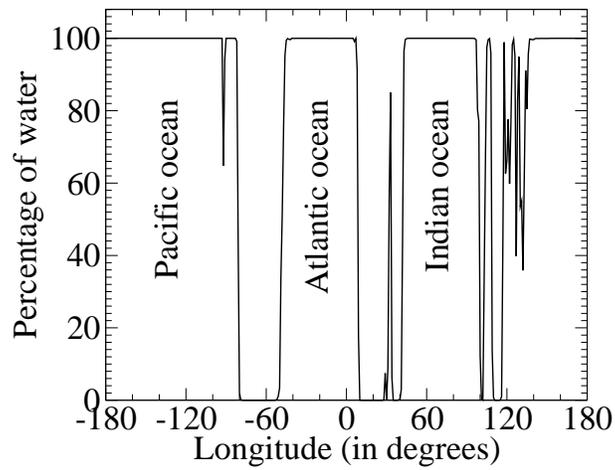}}
\vspace*{0.5cm}
\caption{The percentage of surface covered by water along
         the equator of the Earth. This curve is derived
         from data of the Clouds and the Earth's Radiant 
         Energy System (CERES) instrument on 
         NASA's Earth Observing System (EOS).
         The longitude of 0$^\circ$ corresponds to the
         meridian of Greenwich.
         }
\label{fig15}
\end{figure}

Here, we will show a few examples of the diurnal variation
of the flux and in particular the degree of polarization of a
quasi horizontally inhomogeneous exoplanet
with a longitudinal distribution of land and water similar
to that along the Earth's equator (see Fig.~\ref{fig15}).
For simplicity, we assume the distribution is {\em latitude} 
independent (we thus have a planet with "vertical stripes"). 
Furthermore, the distant observer is located in 
the planet's equatorial plane, and we assume the planet's
obliquity equals zero. Our model planets are somewhat
simpler than those of \citet{2001Natur.412..885F} and 
\citet{2006AsBio...6..881T}, who use
realistically horizontally inhomogeneous planets, but we
do use multiple scattering, like \citet{2006AsBio...6..881T}, 
and polarization.

\begin{figure}
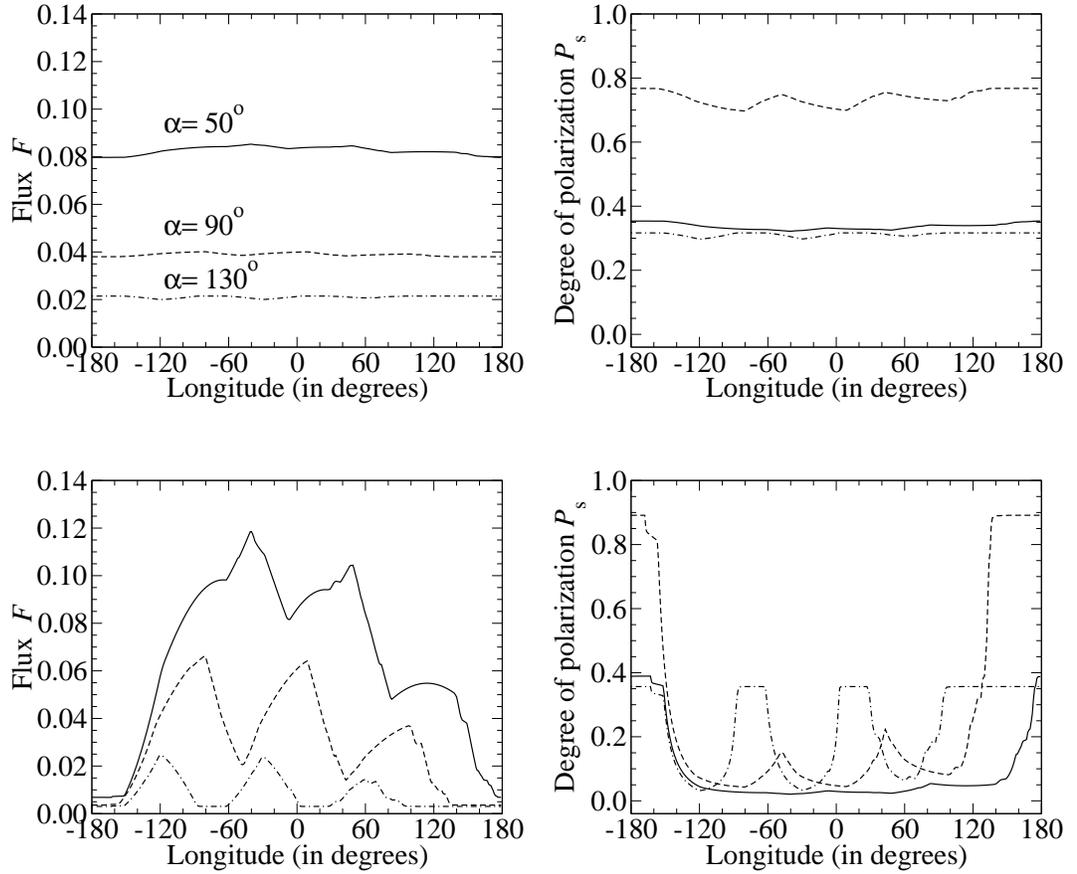

\vspace*{1.0cm}
\centering
\resizebox{14cm}{!}{\includegraphics{fig16a.eps} \hspace*{1.0cm}
                    \includegraphics{fig16b.eps}}
\resizebox{14cm}{!}{\includegraphics{fig16c.eps} \hspace*{1.0cm}
		    \includegraphics{fig16d.eps}}
\vspace*{0.5cm}
\caption{The flux $F$ (left) and degree of polarization $P_{\rm s}$
         (right) of starlight reflected by a cloudless quasi horizontally 
	 inhomogeneous planet as functions of the sub-observer longitude.
	 The longitudinal distribution of land (covered by vegetation)
	 and water (ocean) is as given in Fig.~\ref{fig15}. The 
	 planetary phase angles are 50$^\circ$ (solid lines),
	 90$^\circ$ (dashed lines) and 130$^\circ$ (dash-dotted lines).
	 The upper two panels are for $\lambda=0.44~\mu$m, and the 
	 lower two panels for $\lambda=0.87~\mu$m.
         }
\label{fig16}
\end{figure}

\begin{figure}
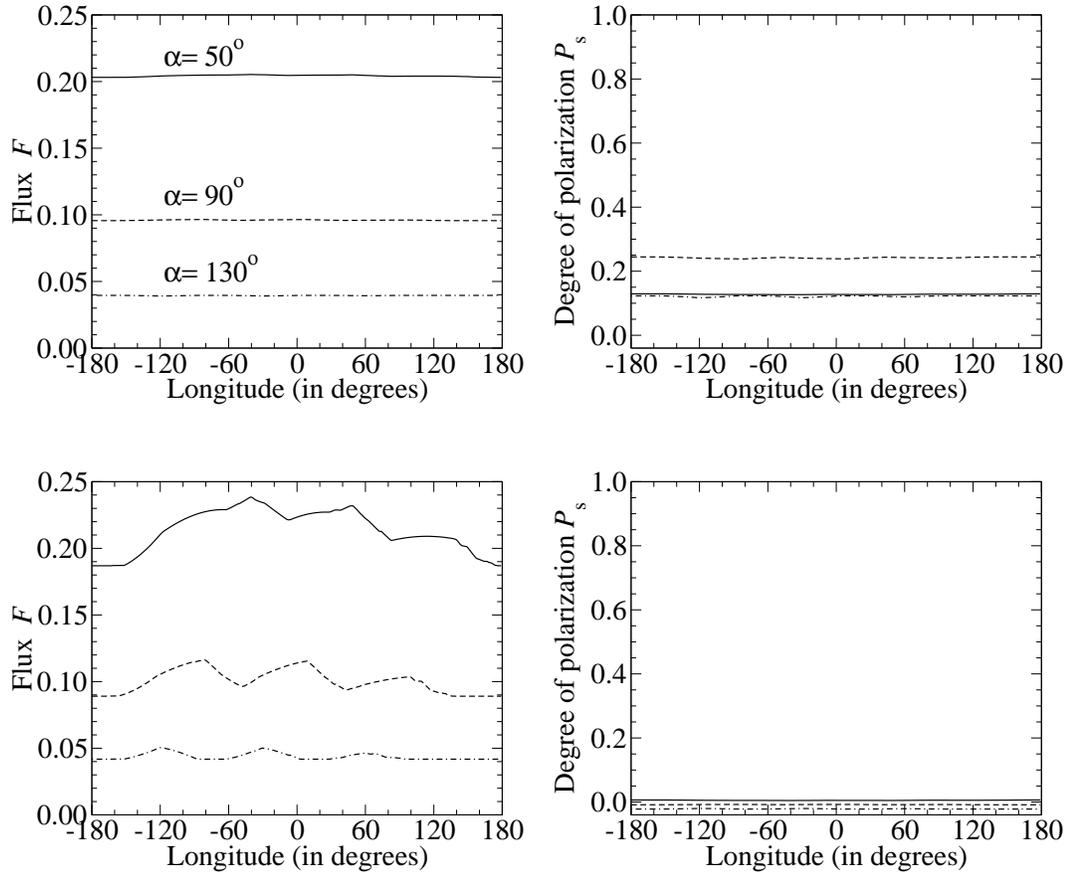

\vspace*{1.0cm}
\centering
\resizebox{14cm}{!}{\includegraphics{fig17a.eps} \hspace*{1.0cm}
                    \includegraphics{fig17b.eps}}
\resizebox{14cm}{!}{\includegraphics{fig17c.eps} \hspace*{1.0cm}
		    \includegraphics{fig17d.eps}}
\vspace*{0.5cm}
\caption{Similar to Fig.~\ref{fig16}, except for planets with a cloud
         coverage of 0.67 \citep{Rossow1996}. The
	 planetary phase angles are 50$^\circ$ (solid lines),
	 90$^\circ$ (dashed lines) and 130$^\circ$ (dash-dotted lines).
	 The upper two panels are for $\lambda=0.44~\mu$m, and the
	 lower two panels for $\lambda=0.87~\mu$m.
         }
\label{fig17}
\end{figure}

In Fig.~\ref{fig16}, we have plotted $F$ and $P_{\rm s}$
of a cloudless planet seen under phase
angles of 50$^\circ$ (when more than half of the illuminated part of
the planetary disk is visible), 90$^\circ$ (quadrature, when half of 
the illuminated planetary disk is visible) and 130$^\circ$ (when 
less than half of the illuminated planetary disk is visible). 
We show curves
for $\lambda=0.44~\mu$m, and for $\lambda=0.87~\mu$m. 
Figure~\ref{fig17}
is similar to Fig.~\ref{fig16}, except that here, like on
an average Earth \citep{Rossow1996}, two-thirds of 
the planet is covered by clouds. 

For the cloudless planet (Fig.~\ref{fig16}), $F$ and $P_{\rm s}$
show for all three phase angles much more variation in the 
near-infrared (0.87~$\mu$m)
than in the blue (0.44~$\mu$m). This is not surprising, 
since in the blue, the vegetation and the ocean are both
dark, and $F$ and $P_{\rm s}$ are thus mostly determined
by the gaseous atmosphere. In the near-infrared, the contrast
between the (red edge of the) vegetation and the ocean shows 
up clearly in both $F$ and $P_{\rm s}$. A similar wavelength 
dependence of the
variation of the reflected flux (at $\alpha=90^\circ$) was found by 
\citet{2001Natur.412..885F} and by \citet{2006AsBio...6..881T}
(the latter have 0.50~to 0.55~$\mu$m as the shortest
wavelength region).
In the blue (0.44~$\mu$m in our plots, and 0.45~$\mu$m in 
the plots by \citet{2001Natur.412..885F}), 
our reflected flux (for $\alpha=90^\circ$) 
shows far less diurnal variation than that of \citet{2001Natur.412..885F}.
This is most likely due to the multiple scattering that we include
in our calculations.
In the infrared (0.87~$\mu$m in our plots, and 0.75~$\mu$m in the 
plots by \citet{2001Natur.412..885F}), 
our maximum fluxes (for $\alpha=90^\circ$)
are larger than those of \citet{2001Natur.412..885F}, namely about 
$0.06$ versus about $0.03$, which cannot be attributed to multiple
scattering, because the atmospheric molecular scattering 
optical thickness at these wavelengths is very small.
This difference in maximum flux is most probably due to 
the difference in surface coverage
(vegetation versus Sahara sand and ocean) and
hence in surface albedo at these wavelengths. 

%The diurnal variation of the reflected flux is expected to
%vary with the planetary phase angle: the larger the angle,
%the smaller the illuminated and observable fraction of the planet,
%and the more uniform the observed surface coverage will be. 
%In the blue ($\lambda=0.44~\mu$m), the Rayleigh scattered light 
%smothers virtually all variation of the flux with phase angle.
%Both for $\alpha=50^\circ$ and for $\alpha=130^\circ$, the flux 
%varies with about 7$\%$ around its average value.
%In the red ($\lambda=0.87~\mu$m), the flux variation with $\alpha$ is 
%stronger; when $\alpha=50^\circ$, the flux varies with 118$~\%$
%around its average value,
%while when $\alpha=130^\circ$, the variation is more than 200$\%$.
 
The diurnal variation in the degree of polarization, 
$P_{\rm s}$, has not been studied before.
In the blue, $P_{\rm s}$ is strongly determined by Rayleigh
scattering, and thus depends strongly on the planetary phase angle.
Here, the average $P_{\rm s}$ for $\alpha=50^\circ$ is very 
similar to that for $\alpha=130^\circ$, because of the 
symmetry of the degree of polarization of light that is singly 
scattered by gaseous molecules with the single scattering angle 
(see Fig.~\ref{fig3}b and Fig.~\ref{fig11}b). The variation 
in $P_{\rm s}$ is largest
(in absolute sense) for $\alpha=90^\circ$, and can be traced
back to the difference, at $\alpha=90^\circ$, between the curves 
for the vegetation--covered
and the ocean--covered planets in Fig.~\ref{fig11}b.
Furthermore according to the difference between these two curves 
in Fig.~\ref{fig11}b, 
the diurnal variation for $\alpha=50^\circ$ should be significantly
larger than that for $\alpha=130^\circ$. 
However, as can be seen in Fig.~\ref{fig16}, the diurnal variation for 
these two phase angles is similar, and even slightly larger for
$\alpha=130^\circ$ than for $\alpha=50^\circ$. 
The reason for this is that when $\alpha=130^\circ$, the illuminated
and observable part of the planet is smaller than for 
$\alpha=50^\circ$, and will therefore have a more uniform surface
coverage (either ocean or vegetation) at any given time. 

At red wavelengths (lower panels in Fig.~\ref{fig16}), where the
largest contribution to the reflected light is from the surface,
the diurnal variation of the degree of polarization, $P_{\rm s}$, 
depends strongly on the planetary phase angle; the variation 
in $P_{\rm s}$ clearly increases with the phase angle.

Next, we discuss the influence of clouds on the diurnal variation
of $F$ and $P_{\rm s}$.
As can be seen in Fig.~\ref{fig17}, and as can be expected from 
Fig.~\ref{fig11}, in the blue (0.44~$\mu$m),
our clouds, which cover two-thirds of the planet,
smother virtually any diurnal variation in $F$ and in $P_{\rm s}$. 
In the near-infrared (0.87~$\mu$m), the flux $F$ retains some
of the diurnal variation, because at this long wavelength 
the vegetation has a much
higher albedo than the ocean (cf. Fig.~\ref{fig12}). 
The fluxes we calculate for the cloudy planets (and $\alpha=90^\circ$)
are very similar to those calculated by \citet{2001Natur.412..885F},
because to describe the light reflected by the cloudy regions on
their model planets, \citet{2001Natur.412..885F} apparently 
do not rely on the single scattering approach they use for the cloud-free
planets. Instead they adopt the reflected fluxes presented
by \citet{1989A&A...214..391H} for cloudy planet models, 
the calculation of which does fully include multiple scattering
(though no vegetated surfaces underneath the clouds).
Qualitatively, our variation in $F$ is very similar to that 
shown by \citet{2006AsBio...6..881T}, although the normalization
is different.

The degree of polarization 
in the near-infrared shows no measurable variation; the
higher albedo of the vegetation only adds more unpolarized
light to the total reflected light.
All clouds on our model planet are optically thick
($b^{\rm c}=10$ at 0.55~$\mu$m, see 
Sect.~\ref{section_modelplanets}). Optically 
thinner clouds, that depolarize the light less, would
probably leave more diurnal variation in $P_{\rm s}$,
as would e.g. polarizing vegetation surfaces 
\citep[][]{2007AAS...210.0906W}.

With our method for calculating the light reflected
by horizontally inhomogeneous planets 
(i.e. using Eq.~\ref{eq_weightedsum}), light reflected by 
e.g. the cloud layer effectively comes from all locations on 
the illuminated
and observed part of the planet, instead of from a limited number
of locations spread over the planet, which would be a more
realistic cloud deck. We might therefore overestimate the 
contribution of the cloud layer to the total reflected signal.
The difference between the signals of our quasi inhomogeneous 
planets and those of more realistically inhomogeneous planets
require more investigation, also with the view on the retrieval
of planetary characteristics.
Such an investigation should also address the region-to-region variations
in time averaged cloud coverage and cloud optical thickness, that are
found on Earth when cloud variations on time scales less than a
month (i.e. weather) are removed 
\citep[for cloud data, see][]{2004BAMS...85..167R}.
Although the actual spatial distribution of cloud properties
across an extrasolar planet might not resemble that of the Earth,
relations between land and cloud coverage.
Regarding flux calculations, examples of region-to-region
variations in cloud coverage are given by 
\citet{2001Natur.412..885F}.

%------------------------------------------------------------------
% Summary
%------------------------------------------------------------------
\section{Summary and conclusion}
\label{section_summary}

We have presented numerical simulations of the flux and state
(degree and direction) of polarization of starlight reflected
by various types of Earth-like extrasolar planets as functions
of the wavelength and as functions of the planetary phase angle.
Spectral fluxes of Earth-like extrasolar planets have been presented
before \citep[see, e.g.][and references therein]{2006AsBio...6...34T,2006ApJ...644L.129T,2006ApJ...651..544M,2006ApJ...644..551T}, 
the spectral variation of the degree 
of polarization of such planets not. 

Our results clearly show that the light reflected by an Earth-like 
exoplanet can be highly polarized, with the degree and direction of 
polarization depending on the physical characteristics of the planetary
atmosphere and surface, on the illumination and viewing geometries, 
and on the wavelength. Polarimetry thus appears to be a useful tool
to distinguish the (polarized) stellar light that has been reflected by
an Earth-like extrasolar planet from the (unpolarized) direct stellar 
light, and hence to detect Earth-like extrasolar planets.

Our results also show that for given physical properties
of the Earth-like extrasolar planet, the degree of polarization of the 
reflected light has similar spectral features as the flux of 
this light: there are features related to the surface albedo of
the planet and high spectral resolution features 
that are due to absorption of light by atmospheric gases.
The latter features, although due to different gases, 
were also present in calculated polarization spectra of
starlight reflected by extrasolar gaseous planets 
\citep[][]{2004A&A...428..663S}. For Earth-like extrasolar 
planets, the occurence of spectral features due to gaseous
absorption in polarization spectra is especially interesting
because these features are conserved upon transmission of the
light through the Earth's atmosphere even if it has a similar 
composition as the extrasolar planet's atmosphere
(although the number of available photons
in the absorption bands will be greatly reduced upon 
traveling through the Earth's atmosphere if it contains the 
same absorbing gases as the extrasolar planet's atmosphere).

Interestingly, the degree of polarization 
of reflected light appears to have a different sensitivity to 
the structure and
composition of the planetary atmosphere and the albedo of 
the underlying surface than the flux of the reflected light has.
Compared to using only spectroscopy, polarimetry could thus provide 
additional and different information
on the structure and composition of an extrasolar planet's atmosphere
and surface.
In particular, we have shown that the 
degree of polarization at continuum wavelengths around
the O$_2$ $A$-band (around 0.76 nm) is sensitive to the 
cloud top altitude, whereas the continuum flux in this 
wavelength region is virtually
insensitive to the cloud top altitude.
The degree of polarization could thus be used for cloud top
altitude determination, a method that is also applied 
to e.g. observations of the Earth observing POLDER instrument 
\citep[][]{1994Deschamps} and its successors.
We have also shown that the depth of the O$_2$ $A$-band in a 
polarization spectrum is sensitive to the oxygen mixing ratio,
whereas the flux in the O$_2$ $A$-band is sensitive to
both the cloud top altitude and the oxygen mixing ratio.
Polarimetry could thus help to disentangle crucial information
on gaseous mixing ratios and clouds from the sparse, spatially
unresolved data that will be available for exoplanets.

Because of the strengths for detecting and characterizing 
Earth-like exoplanets, as mentioned above, polarimetry is a 
technique used in 
SEE-COAST (the Super Earths Explorer -- Coronographic Off-Axis 
Telescope), a space-based telescope for the detection and the 
characterization of gaseous exoplanets and large rocky exoplanets, 
so called 'Super-Earths', \citep[][]{2006sf2a.conf..429S}, that 
has been proposed to ESA in response to its 2007 Cosmic Vision call.

Our numerical simulations should not only be useful
for researchers that are interested in designing and building 
(spectro)polarimeters
for the detection and characterization of Earth-like exoplanets, 
and for researchers interested in measuring and analyzing 
the state of polarization of extrasolar planets. 
Indeed, due to their optics,
spectrometers tend to be sensitive to the state of polarization 
of the incoming light (unless carefully corrected for): the
measured fluxes thus depend on the degree and direction of
polarization of the observed light \citep[see e.g.][]{2000JGR...10522379S}. 
In case the polarization sensitivity of a spectrometer is known,
e.g. because it has been measured during calibration, our
numerical simulations can help to {\em estimate} the error in the 
measured flux. To actually {\em correct} measured fluxes for the
polarization sensitivity of an instrument, one has to know the
instrument's polarization sensitivity {\em and} one has to measure the
state of polarization of the incoming light, because as our
simulations show, this state of polarization varies with
the illumination and viewing geometries and with the physical
characteristics of the planet. Because the state of polarization
is wavelength dependent and shows high spectral resolution
features similar to those in the total flux, the state of 
polarization should be measured with the same spectral
resolution as the total flux if one is interested in accurately 
measuring the depth and shape of e.g. gaseous absorption bands.

Finally, \citet{2005A&A...444..275S} showed that neglecting polarization
when calculating disk--integrated total flux spectra of gaseous planets 
induces errors of several percent across the continuum and gaseous 
absorption bands. The reason for these errors is that the flux of
scattered light depends on the state of polarization of the incident
light \citep[see also][]{1994JQSRT..51..491M,1998GeoRL..25..135L},
and while the incident stellar light can be assumed to be unpolarized
\citep[][]{1987Natur.326..270K}, it is usually polarized upon the first
scattering within the atmosphere. Hence, as long as only single
scattering processes are taken into account, polarization does
not influence the scattered total flux, and can thus safely be ignored
if one is only interested in total fluxes. When multiple
scattering has to be accounted for, however, neglecting polarization
and treating light as a scalar instead of as a vector,
does influence the scattered total flux.
The fluxes presented in this paper have all been calculated taking
polarization fully into account. To investigate the influence of 
neglecting polarization on the fluxes of Earth-like extrasolar
planets, we performed a number of calculations without polarization.

\begin{figure}
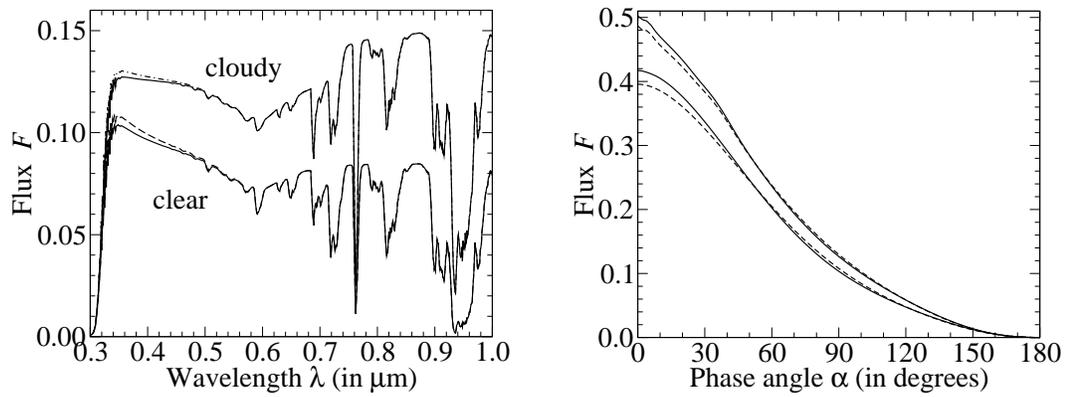

\vspace*{1.0cm}
\centering
\resizebox{14cm}{!}{\includegraphics{fig18a.eps} \hspace*{2.5cm}
                    \includegraphics{fig18b.eps} }
\vspace*{0.5cm}
\vspace*{0.5cm}
\caption{Total reflected fluxes as functions of the
         wavelength for $\alpha=90^\circ$ (left), 
         and as functions
         of the phase angle for $\lambda=0.35~\mu$m (right). 
         The solid lines have been calculated with polarization, 
         and the dashed lines without polarization. 
         The lower two lines are for a homogeneous, cloudfree planet
         with a surface albedo of 0.4, and the upper two lines are
         for a homogeneous, cloudy planet with a surface covered
         by vegetation.}
\label{fig18}
\end{figure}

Figure~\ref{fig18} shows total fluxes calculated with and without
polarization, for a cloudfree and a cloudy planet, as functions
of the wavelength at quadrature,
and at $\lambda=0.35~\mu$m as functions of the phase angle.
For a cloudfree atmosphere with the surface albedo of 0.4, the 
maximum error is about 4$\%$, at a wavelength of 0.35~$\mu$m.
For the cloudy atmosphere with the vegetation underneath, the
maximum error is about 2.5~$\%$, also near 0.35~$\mu$m.
At other phase angles (Fig.~\ref{fig18}b), 
the relative errors have similar sizes, although they vanish 
near $\alpha=50^\circ$ and 120$^\circ$.
The errors have a similar phase angle dependence, but 
are smaller than those presented for gaseous planets by
\citet{2005A&A...444..275S}. The errors are smaller 
because our Earth-like atmosphere
is optically much thinner than that of a gaseous planet, and
hence less multiple scattering takes place. That for an
Earth-like planet (with a total molecular scattering
optical thickness of about 0.1 at $\lambda=0.55~\mu$m, see
Table~\ref{tab1}), the errors will be smaller is also clear
from Fig.~5 of \citet{2005A&A...444..275S}, which shows
the errors due to neglecting polarization in planetary albedos
as functions of the atmospheric molecular scattering optical 
thickness. 
The decrease of the error with decreasing molecular scattering 
optical thickness also explains the decrease of the error with
wavelength in Fig.~\ref{fig18}, and the smaller error for the
cloudy atmosphere.
Concluding, for Earth-like model atmospheres, neglecting
polarization when calculating fluxes leads to maximum errors of 
a few percent in the continuum.

%-----------------------------------------------------------------------
% References
%-----------------------------------------------------------------------
\bibliographystyle{aa} % style aa.bst
\bibliography{refs_astroph}

\begin{thebibliography}{80}
\expandafter\ifx\csname natexlab\endcsname\relax\def\natexlab#1{#1}\fi

\bibitem[{{Aben} {et~al.}(1997){Aben}, {Helderman}, {Stam}, \&
  {Stammes}}]{1997ABEN}
{Aben}, I., {Helderman}, F., {Stam}, D., \& {Stammes}, P. 1997, in
  Polarization: Measurement, Analysis, and Remote Sensing. Proceedings SPIE
  {\bf 3121}, ed. D.~{Goldstein} \& R.~{Chipman}, 446--451

\bibitem[{{Aben} {et~al.}(1999){Aben}, {Helderman}, {Stam}, \&
  {Stammes}}]{1999GeoRL..26..591A}
{Aben}, I., {Helderman}, F., {Stam}, D.~M., \& {Stammes}, P. 1999, Geophys.
  Res. Lett., 26, 591

\bibitem[{{Aben} {et~al.}(2001){Aben}, {Stam}, \&
  {Helderman}}]{2001GeoRL..28..519A}
{Aben}, I., {Stam}, D.~M., \& {Helderman}, F. 2001, Geophys. Res. Lett., 28,
  519

\bibitem[{{Arnold} {et~al.}(2002){Arnold}, {Gillet}, {Lardi{\`e}re}, {Riaud},
  \& {Schneider}}]{2002A&A...392..231A}
{Arnold}, L., {Gillet}, S., {Lardi{\`e}re}, O., {Riaud}, P., \& {Schneider}, J.
  2002, Astron. \& Astrophys., 392, 231

\bibitem[{{Bailey}(2007)}]{2007AsBio...7..320B}
{Bailey}, J. 2007, Astrobiology, 7, 320

\bibitem[{{Bates}(1984)}]{1984P&SS...32..785B}
{Bates}, D.~R. 1984, Planetary Space Scie., 32, 785

\bibitem[{{Beuzit} {et~al.}(2006){Beuzit}, {Feldt}, {Dohlen}, {Mouillet},
  {Puget}, {Antici}, {Baruffolo}, {Baudoz}, {Berton}, {Boccaletti},
  {Carbillet}, {Charton}, {Claudi}, {Downing}, {Feautrier}, {Fedrigo}, {Fusco},
  {Gratton}, {Hubin}, {Kasper}, {Langlois}, {Moutou}, {Mugnier}, {Pragt},
  {Rabou}, {Saisse}, {Schmid}, {Stadler}, {Turrato}, {Udry}, {Waters}, \&
  {Wildi}}]{2006Msngr.125...29B}
{Beuzit}, J.-L., {Feldt}, M., {Dohlen}, K., {et~al.} 2006, The Messenger, 125,
  29

\bibitem[{{Br{\'e}on} \& {Henriot}(2006)}]{2006JGRC..11106005B}
{Br{\'e}on}, F.~M. \& {Henriot}, N. 2006, Journal of Geophysical Research
  (Oceans), 111, 6005

\bibitem[{{Chowdhary} {et~al.}(2002){Chowdhary}, {Cairns}, \&
  {Travis}}]{2002JAtS...59..383C}
{Chowdhary}, J., {Cairns}, B., \& {Travis}, L.~D. 2002, Journal of Atmospheric
  Sciences, 59, 383

\bibitem[{{Cox} \& {Munk}(1954)}]{1954JOSA...44..838C}
{Cox}, C. \& {Munk}, W. 1954, Journal of the Optical Society of America
  (1917-1983), 44, 838

\bibitem[{{de Haan} {et~al.}(1987){de Haan}, {Bosma}, \&
  {Hovenier}}]{1987A&A...183..371D}
{de Haan}, J.~F., {Bosma}, P.~B., \& {Hovenier}, J.~W. 1987, Astron. \&
  Astrophys., 183, 371

\bibitem[{{de Rooij} \& {van der Stap}(1984)}]{1984A&A...131..237D}
{de Rooij}, W.~A. \& {van der Stap}, C.~C.~A.~H. 1984, Astron. \& Astrophys.,
  131, 237

\bibitem[{{Deschamps} {et~al.}(1994){Deschamps}, {Br\'{e}on}, {Leroy},
  {Podaire}, {Bricaud}, {Buriez}, \& {Seze}}]{1994Deschamps}
{Deschamps}, P.~Y., {Br\'{e}on}, F.~M., {Leroy}, M., {et~al.} 1994, IEEE Trans.
  Geosci. Remote Sens., 32, 598

\bibitem[{{Fischer} {et~al.}(1991){Fischer}, {Cordes}, {Schmitz-Peiffer},
  {Renger}, \& {M{\"o}rl}}]{1991JApMe..30.1260F}
{Fischer}, J., {Cordes}, W., {Schmitz-Peiffer}, A., {Renger}, W., \&
  {M{\"o}rl}, P. 1991, Journal of Applied Meteorology, 30, 1260

\bibitem[{{Fischer} \& {Grassl}(1991)}]{1991JApMe..30.1245F}
{Fischer}, J. \& {Grassl}, H. 1991, Journal of Applied Meteorology, 30, 1245

\bibitem[{{Ford} {et~al.}(2001){Ford}, {Seager}, \&
  {Turner}}]{2001Natur.412..885F}
{Ford}, E.~B., {Seager}, S., \& {Turner}, E.~L. 2001, Nature, 412, 885

\bibitem[{{Gandorfer} {et~al.}(2004){Gandorfer}, {Steiner}, {Aebersold},
  {Egger}, {Feller}, {Gisler}, {Hagenbuch}, \& {Stenflo}}]{2004A&A...422..703G}
{Gandorfer}, A.~M., {Steiner}, H.~P.~P.~P., {Aebersold}, F., {et~al.} 2004,
  Astron. \& Astrophys., 422, 703

\bibitem[{{Gisler} {et~al.}(2004){Gisler}, {Schmid}, {Thalmann}, {Povel},
  {Stenflo}, {Joos}, {Feldt}, {Lenzen}, {Tinbergen}, {Gratton}, {Stuik},
  {Stam}, {Brandner}, {Hippler}, {Turatto}, {Neuhauser}, {Dominik}, {Hatzes},
  {Henning}, {Lima}, {Quirrenbach}, {Waters}, {Wuchterl}, \&
  {Zinnecker}}]{2004SPIE.5492..463G}
{Gisler}, D., {Schmid}, H.~M., {Thalmann}, C., {et~al.} 2004, in Presented at
  the Society of Photo-Optical Instrumentation Engineers (SPIE) Conference,
  Vol. 5492, Ground-based Instrumentation for Astronomy. Edited by Alan F. M.
  Moorwood and Iye Masanori. Proceedings of the SPIE, Volume 5492, pp. 463-474
  (2004)., ed. A.~F.~M. {Moorwood} \& M.~{Iye}, 463--474

\bibitem[{{Grainger} \& {Ring}(1962)}]{1962Natur.193..762W}
{Grainger}, J.~F. \& {Ring}, J. 1962, Nature, 193, 762

\bibitem[{{Haferman} {et~al.}(1997){Haferman}, {Smith}, \&
  {Krajewski}}]{1997Haferman}
{Haferman}, J.~L., {Smith}, T.~F., \& {Krajewski}, W.~F. 1997, Journal of
  Quantitative Spectroscopy and Radiative Transfer, 58, 379

\bibitem[{{Hamdani} {et~al.}(2006){Hamdani}, {Arnold}, {Foellmi}, {Berthier},
  {Billeres}, {Briot}, {Fran{\c c}ois}, {Riaud}, \&
  {Schneider}}]{2006A&A...460..617H}
{Hamdani}, S., {Arnold}, L., {Foellmi}, C., {et~al.} 2006, Astron. \&
  Astrophys., 460, 617

\bibitem[{{Hansen} \& {Hovenier}(1974{\natexlab{a}})}]{1974JAtS...31.1137H}
{Hansen}, J.~E. \& {Hovenier}, J.~W. 1974{\natexlab{a}}, Journal of Atmospheric
  Sciences, 31, 1137

\bibitem[{{Hansen} \& {Hovenier}(1974{\natexlab{b}})}]{1974IAUS...65..197H}
{Hansen}, J.~E. \& {Hovenier}, J.~W. 1974{\natexlab{b}}, in IAU Symp. 65:
  Exploration of the Planetary System, 197--200

\bibitem[{{Hansen} \& {Travis}(1974)}]{1974SSRv...16..527H}
{Hansen}, J.~E. \& {Travis}, L.~D. 1974, Space Science Reviews, 16, 527

\bibitem[{{Hough} \& {Lucas}(2003)}]{2003toed.conf...11H}
{Hough}, J.~H. \& {Lucas}, P.~W. 2003, in ESA SP-539: Earths: DARWIN/TPF and
  the Search for Extrasolar Terrestrial Planets, 11--17

\bibitem[{{Hough} {et~al.}(2006{\natexlab{a}}){Hough}, {Lucas}, {Bailey},
  {Tamura}, \& {Hirst}}]{2006SPIE.6269E..25H}
{Hough}, J.~H., {Lucas}, P.~W., {Bailey}, J.~A., {Tamura}, M., \& {Hirst}, E.
  2006{\natexlab{a}}, in the Society of Photo-Optical Instrumentation Engineers
  (SPIE) Conference, Vol. 6269, Ground-based and Airborne Instrumentation for
  Astronomy. Edited by McLean, Ian S.; Iye, Masanori. Proceedings of the SPIE,
  Volume 6269, pp. 62690S (2006).

\bibitem[{{Hough} {et~al.}(2006{\natexlab{b}}){Hough}, {Lucas}, {Bailey},
  {Tamura}, {Hirst}, {Harrison}, \& {Bartholomew-Biggs}}]{2006PASP..118.1305H}
{Hough}, J.~H., {Lucas}, P.~W., {Bailey}, J.~A., {et~al.} 2006{\natexlab{b}},
  Pub. Astron. Soc. Pac., 118, 1305

\bibitem[{{Hovenier} \& {Hage}(1989)}]{1989A&A...214..391H}
{Hovenier}, J.~W. \& {Hage}, J.~I. 1989, Astron. \& Astrophys., 214, 391

\bibitem[{{Hovenier} {et~al.}(2004){Hovenier}, {van der Mee}, \&
  {Domke}}]{2004Hovenier}
{Hovenier}, J.~W., {van der Mee}, C., \& {Domke}, H. 2004, {Transfer of
  Polarized Light in Planetary Atmospheres; Basic Concepts and Practical
  Methods} (Kluwer, Dordrecht; Springer, Berlin)

\bibitem[{{Joos} \& {Schmid}(2007)}]{2007A&A...463.1201J}
{Joos}, F. \& {Schmid}, H.~M. 2007, Astron. \& Astrophys., 463, 1201

\bibitem[{{Joos} {et~al.}(2005){Joos}, {Schmid}, {Gisler}, {Feldt}, {Brandner},
  {Stam}, {Quirrenbach}, \& {Stuik}}]{2005ASPC..343..189J}
{Joos}, F., {Schmid}, H.~M., {Gisler}, D., {et~al.} 2005, in Astronomical
  Society of the Pacific Conference Series, ed. A.~{Adamson}, C.~{Aspin}, \&
  C.~{Davis}, 189--+

\bibitem[{{Keller}(2006)}]{2006SPIE.6269E..26K}
{Keller}, C.~U. 2006, in Presented at the Society of Photo-Optical
  Instrumentation Engineers (SPIE) Conference, Vol. 6269, Ground-based and
  Airborne Instrumentation for Astronomy. Edited by McLean, Ian S.; Iye,
  Masanori. Proceedings of the SPIE, Volume 6269, pp. 62690T (2006).

\bibitem[{{Kemp} {et~al.}(1987){Kemp}, {Henson}, {Steiner}, \&
  {Powell}}]{1987Natur.326..270K}
{Kemp}, J.~C., {Henson}, G.~D., {Steiner}, C.~T., \& {Powell}, E.~R. 1987,
  Nature, 326, 270

\bibitem[{{Kiang} {et~al.}(2007){Kiang}, {Segura}, {Tinetti}, {Govindjee},
  {Blankenship}, {Cohen}, {Siefert}, {Crisp}, \&
  {Meadows}}]{2007AsBio...7..252K}
{Kiang}, N.~Y., {Segura}, A., {Tinetti}, G., {et~al.} 2007, Astrobiology, 7,
  252

\bibitem[{{Koepke}(1984)}]{1984ApOpt..23.1816K}
{Koepke}, P. 1984, Appl. Optics, 23, 1816

\bibitem[{{Kuze} \& {Chance}(1994)}]{1994JGR....9914481K}
{Kuze}, A. \& {Chance}, K.~V. 1994, J. Geophys. Res., 99, 14481

\bibitem[{{Lacis} {et~al.}(1998){Lacis}, {Chowdhary}, {Mishchenko}, \&
  {Cairns}}]{1998GeoRL..25..135L}
{Lacis}, A.~A., {Chowdhary}, J., {Mishchenko}, M.~I., \& {Cairns}, B. 1998,
  Geophys. Res. Lett., 25, 135

\bibitem[{{Lacis} \& {Oinas}(1991)}]{1991JGR....96.9027L}
{Lacis}, A.~A. \& {Oinas}, V. 1991, J. Geophys. Res., 96, 9027

\bibitem[{{Liou} \& {Takano}(2002)}]{2002GeoRL..29i..27L}
{Liou}, K.~N. \& {Takano}, Y. 2002, Geophys. Res. Lett., 29, 27

\bibitem[{{McClatchey} {et~al.}(1972){McClatchey}, {Fenn}, {Selby}, {Volz}, \&
  {Garing}}]{1972McClatchey}
{McClatchey}, R., {Fenn}, R., {Selby}, J., {Volz}, F., \& {Garing}, J. 1972,
  {Optical Properties of the Atmosphere, AFCRL-72.0497} (U.S. Air Force
  Cambridge Research Labs)

\bibitem[{{Mishchenko} {et~al.}(1994){Mishchenko}, {Lacis}, \&
  {Travis}}]{1994JQSRT..51..491M}
{Mishchenko}, M.~I., {Lacis}, A.~A., \& {Travis}, L.~D. 1994, Journal of
  Quantitative Spectroscopy and Radiative Transfer, 51, 491

\bibitem[{{Monta{\~n}{\'e}s-Rodr{\'{\i}}guez}
  {et~al.}(2006){Monta{\~n}{\'e}s-Rodr{\'{\i}}guez}, {Pall{\'e}}, {Goode}, \&
  {Mart{\'{\i}}n-Torres}}]{2006ApJ...651..544M}
{Monta{\~n}{\'e}s-Rodr{\'{\i}}guez}, P., {Pall{\'e}}, E., {Goode}, P.~R., \&
  {Mart{\'{\i}}n-Torres}, F.~J. 2006, Astrophys. J., 651, 544

\bibitem[{{Peck} \& {Reeder}(1972)}]{1972JOSA...62..958P}
{Peck}, E.~R. \& {Reeder}, K. 1972, Journal of the Optical Society of America
  (1917-1983), 62, 958

\bibitem[{{Preusker} {et~al.}(1995){Preusker}, {B\"{o}ttger}, \&
  {Fischer}}]{1995PREUSKER}
{Preusker}, R., {B\"{o}ttger}, U., \& {Fischer}, J. 1995, in Atmospheric
  Sensing and Modeling II. Proceedings SPIE 2582, ed. R.~{Santer}, 13--20

\bibitem[{{Rossow} \& {Due{\~n}as}(2004)}]{2004BAMS...85..167R}
{Rossow}, W.~B. \& {Due{\~n}as}, E.~N. 2004, Bulletin of the American
  Meteorological Society, vol.~85, Issue 2, pp.167-172, 85, 167

\bibitem[{{Rossow} {et~al.}(1996){Rossow}, {Walker}, {Beuschel}, \&
  {Roiter}}]{Rossow1996}
{Rossow}, W.~B., {Walker}, A.~W., {Beuschel}, D.~E., \& {Roiter}, M.~D. 1996,
  {International Satellite Cloud Climatology Project (ISCCP) Documentation of
  New Cloud Datasets. WMO/TD-No. 737} (World Meteorological Organization,
  Geneva, Switzerland)

\bibitem[{{Rothman} {et~al.}(2005){Rothman}, {Jacquemart}, {Barbe}, {Benner},
  {Birk}, {Brown}, {Carleer}, {Chackerian}, {Chance}, {Coudert}, {Dana},
  {Devi}, {Flaud}, {Gamache}, {Goldman}, {Hartmann}, {Jucks}, {Maki}, {Mandin},
  {Massie}, {Orphal}, {Perrin}, {Rinsland}, {Smith}, {Tennyson}, {Tolchenov},
  {Toth}, {Vander Auwera}, {Varanasi}, \& {Wagner}}]{2005JQSRT..96..139R}
{Rothman}, L.~S., {Jacquemart}, D., {Barbe}, A., {et~al.} 2005, Journal of
  Quantitative Spectroscopy and Radiative Transfer, 96, 139

\bibitem[{{Saar} \& {Seager}(2003)}]{2003ASPC..294..529S}
{Saar}, S.~H. \& {Seager}, S. 2003, in Astronomical Society of the Pacific
  Conference Series, 529--534

\bibitem[{{Sagan} {et~al.}(1993){Sagan}, {Thompson}, {Carlson}, {Gurnett}, \&
  {Hord}}]{1993Natur.365..715S}
{Sagan}, C., {Thompson}, W.~R., {Carlson}, R., {Gurnett}, D., \& {Hord}, C.
  1993, Nature, 365, 715

\bibitem[{{Saiedy} {et~al.}(1967){Saiedy}, {Jacobowitz}, \&
  {Wark}}]{1967JAtS...24...63S}
{Saiedy}, F., {Jacobowitz}, H., \& {Wark}, D.~Q. 1967, Journal of Atmospheric
  Sciences, 24, 63

\bibitem[{{Schmid} {et~al.}(2005){Schmid}, {Gisler}, {Joos}, {Povel},
  {Stenflo}, {Feldt}, {Lenzen}, {Brandner}, {Tinbergen}, {Quirrenbach},
  {Stuik}, {Gratton}, {Turatto}, \& {Neuh{\"a}user}}]{2005ASPC..343...89S}
{Schmid}, H.~M., {Gisler}, D., {Joos}, F., {et~al.} 2005, in Astronomical
  Society of the Pacific Conference Series, Vol. 343, Astronomical Polarimetry:
  Current Status and Future Directions, ed. A.~{Adamson}, C.~{Aspin},
  C.~{Davis}, \& T.~{Fujiyoshi}, 89--+

\bibitem[{{Schmid} {et~al.}(2006){Schmid}, {Joos}, \&
  {Tschan}}]{2006A&A...452..657S}
{Schmid}, H.~M., {Joos}, F., \& {Tschan}, D. 2006, Astron. \& Astrophys., 452,
  657

\bibitem[{{Schneider} {et~al.}(2006){Schneider}, {Riaud}, {Tinetti}, {Schmid},
  {Stam}, {Udry}, {Baudoz}, {Boccaletti}, {Grasset}, {Mawet}, {Surdej}, \& {The
  See-Coast Team}}]{2006sf2a.conf..429S}
{Schneider}, J., {Riaud}, P., {Tinetti}, G., {et~al.} 2006, in SF2A-2006:
  Semaine de l'Astrophysique Francaise, ed. D.~{Barret}, F.~{Casoli},
  G.~{Lagache}, A.~{Lecavelier}, \& L.~{Pagani}, 429--+

\bibitem[{{Seager} {et~al.}(2005){Seager}, {Turner}, {Schafer}, \&
  {Ford}}]{2005AsBio...5..372S}
{Seager}, S., {Turner}, E.~L., {Schafer}, J., \& {Ford}, E.~B. 2005,
  Astrobiology, 5, 372

\bibitem[{{Seager} {et~al.}(2000){Seager}, {Whitney}, \&
  {Sasselov}}]{2000ApJ...540..504S}
{Seager}, S., {Whitney}, B.~A., \& {Sasselov}, D.~D. 2000, Astrophys. J., 540,
  504

\bibitem[{{Segura} {et~al.}(2005){Segura}, {Kasting}, {Meadows}, {Cohen},
  {Scalo}, {Crisp}, {Butler}, \& {Tinetti}}]{2005AsBio...5..706S}
{Segura}, A., {Kasting}, J.~F., {Meadows}, V., {et~al.} 2005, Astrobiology, 5,
  706

\bibitem[{{Sengupta} \& {Maiti}(2006)}]{2006ApJ...639.1147S}
{Sengupta}, S. \& {Maiti}, M. 2006, Astrophys. J., 639, 1147

\bibitem[{{Shkuratov} {et~al.}(2005){Shkuratov}, {Kreslavsky}, {Kaydash},
  {Videen}, {Bell}, {Wolff}, {Hubbard}, {Noll}, \&
  {Lubenow}}]{2005Icar..176....1S}
{Shkuratov}, Y., {Kreslavsky}, M., {Kaydash}, V., {et~al.} 2005, Icarus, 176, 1

\bibitem[{{Sromovsky}(2005)}]{2005Icar..173..254S}
{Sromovsky}, L.~A. 2005, Icarus, 173, 254

\bibitem[{{Stam} {et~al.}(2000{\natexlab{a}}){Stam}, {de Haan}, {Hovenier}, \&
  {Stammes}}]{2000JQSRTStam}
{Stam}, D., {de Haan}, J., {Hovenier}, J., \& {Stammes}, P. 2000{\natexlab{a}},
  Journal of Quantitative Spectroscopy and Radiative Transfer, 64, 131

\bibitem[{{Stam} {et~al.}(2005){Stam}, {Hovenier}, \&
  {Waters}}]{2005ASPC..343..207S}
{Stam}, D., {Hovenier}, J., \& {Waters}, L. 2005, in Astronomical Society of
  the Pacific Conference Series, ed. A.~{Adamson}, C.~{Aspin}, \& C.~{Davis},
  207--+

\bibitem[{{Stam}(2003)}]{2003toed.conf..615S}
{Stam}, D.~M. 2003, in ESA SP-539: Earths: DARWIN/TPF and the Search for
  Extrasolar Terrestrial Planets, 615--619

\bibitem[{{Stam} {et~al.}(2002){Stam}, {Aben}, \&
  {Helderman}}]{2002JGRD.107t.AAC1S}
{Stam}, D.~M., {Aben}, I., \& {Helderman}, F. 2002, Journal of Geophysical
  Research (Atmospheres), AAC 1

\bibitem[{{Stam} {et~al.}(2000{\natexlab{b}}){Stam}, {De Haan}, {Hovenier}, \&
  {Aben}}]{2000JGR...10522379S}
{Stam}, D.~M., {De Haan}, J.~F., {Hovenier}, J.~W., \& {Aben}, I.
  2000{\natexlab{b}}, J. Geophys. Res., 22379

\bibitem[{{Stam} {et~al.}(1999){Stam}, {De Haan}, {Hovenier}, \&
  {Stammes}}]{1999JGR...10416843S}
{Stam}, D.~M., {De Haan}, J.~F., {Hovenier}, J.~W., \& {Stammes}, P. 1999, J.
  Geophys. Res., 104, 16843

\bibitem[{{Stam} {et~al.}(2006){Stam}, {de Rooij}, {Cornet}, \&
  {Hovenier}}]{2006A&A...452..669S}
{Stam}, D.~M., {de Rooij}, W.~A., {Cornet}, G., \& {Hovenier}, J.~W. 2006,
  Astron. \& Astrophys., 452, 669

\bibitem[{{Stam} {et~al.}(2003){Stam}, {Hovenier}, \&
  {Waters}}]{2003ASPC..294..535S}
{Stam}, D.~M., {Hovenier}, J., \& {Waters}, R. 2003, in Astronomical Society of
  the Pacific Conference Series, Vol. 294, Scientific Frontiers in Research on
  Extrasolar Planets, ed. D.~{Deming} \& S.~{Seager}, 535--538

\bibitem[{{Stam} \& {Hovenier}(2005)}]{2005A&A...444..275S}
{Stam}, D.~M. \& {Hovenier}, J.~W. 2005, Astron. \& Astrophys., 444, 275

\bibitem[{{Stam} {et~al.}(2004){Stam}, {Hovenier}, \&
  {Waters}}]{2004A&A...428..663S}
{Stam}, D.~M., {Hovenier}, J.~W., \& {Waters}, L.~B.~F.~M. 2004, Astron. \&
  Astrophys., 428, 663

\bibitem[{{Stammes} {et~al.}(1994){Stammes}, {Kuik}, \& {de
  Haan}}]{1994STAMMES}
{Stammes}, P., {Kuik}, F., \& {de Haan}, J. 1994, in Proceedings PIERS 1994,
  Kluwer Acad., Dordrecht, ed. B.~e.~a. {Arbesser-Rastburg}, 2255--2259

\bibitem[{{Tinetti} {et~al.}(2006{\natexlab{a}}){Tinetti}, {Meadows}, {Crisp},
  {Fong}, {Fishbein}, {Turnbull}, \& {Bibring}}]{2006AsBio...6...34T}
{Tinetti}, G., {Meadows}, V.~S., {Crisp}, D., {et~al.} 2006{\natexlab{a}},
  Astrobiology, 6, 34

\bibitem[{{Tinetti} {et~al.}(2006{\natexlab{b}}){Tinetti}, {Meadows}, {Crisp},
  {Kiang}, {Kahn}, {Fishbein}, {Velusamy}, \& {Turnbull}}]{2006AsBio...6..881T}
{Tinetti}, G., {Meadows}, V.~S., {Crisp}, D., {et~al.} 2006{\natexlab{b}},
  Astrobiology, 6, 881

\bibitem[{{Tinetti} {et~al.}(2006{\natexlab{c}}){Tinetti}, {Rashby}, \&
  {Yung}}]{2006ApJ...644L.129T}
{Tinetti}, G., {Rashby}, S., \& {Yung}, Y.~L. 2006{\natexlab{c}}, Astrophys.
  J., 644, L129

\bibitem[{{Turnbull} {et~al.}(2006){Turnbull}, {Traub}, {Jucks}, {Woolf},
  {Meyer}, {Gorlova}, {Skrutskie}, \& {Wilson}}]{2006ApJ...644..551T}
{Turnbull}, M.~C., {Traub}, W.~A., {Jucks}, K.~W., {et~al.} 2006, Astrophys.
  J., 644, 551

\bibitem[{{van de Hulst}(1957)}]{1957vandeHulst}
{van de Hulst}, H.~C. 1957, {Light Scattering by Small Particles} (J. Wiley and
  Sons, New York.)

\bibitem[{{van de Hulst}(1980)}]{1980vandeHulst}
{van de Hulst}, H.~C. 1980, {Multiple Light Scattering, Tables, Formulas, and
  Applications, {\rm Vols. 1 and 2}} (Academic Press, New York.)

\bibitem[{{van Deelen} {et~al.}(2005){van Deelen}, {Landgraf}, \&
  {Aben}}]{2005JQSRT..95..309V}
{van Deelen}, R., {Landgraf}, J., \& {Aben}, I. 2005, Journal of Quantitative
  Spectroscopy and Radiative Transfer, 95, 309

\bibitem[{{Wolstencroft} {et~al.}(2007){Wolstencroft}, {Breon}, \&
  {Tranter}}]{2007AAS...210.0906W}
{Wolstencroft}, R.~D., {Breon}, F., \& {Tranter}, G. 2007, in American
  Astronomical Society Meeting Abstracts, Vol. 210, American Astronomical
  Society Meeting Abstracts, 09.06--+

\bibitem[{{Wolstencroft} \& {Raven}(2002)}]{2002Icar..157..535W}
{Wolstencroft}, R.~D. \& {Raven}, J.~A. 2002, Icarus, 157, 535

\bibitem[{{Woolf} {et~al.}(2002){Woolf}, {Smith}, {Traub}, \&
  {Jucks}}]{2002ApJ...574..430W}
{Woolf}, N.~J., {Smith}, P.~S., {Traub}, W.~A., \& {Jucks}, K.~W. 2002,
  Astrophys. J., 574, 430

\end{thebibliography}

%-----------------------------------------------------------------------
% Figure 1:
%-----------------------------------------------------------------------
\clearpage
\newpage

%-----------------------------------------------------------------------
% Figure 2:
%-----------------------------------------------------------------------
%\clearpage
%\newpage
%
%\begin{figure}
%\vspace*{5.0cm}
%\centering
%\resizebox{8cm}{!}{\includegraphics{fig2.eps}}
%\vspace*{0.5cm}
%\caption{The cumulative probability function for finding an 
%         observable planet at a given phase angle $\alpha$, i.e. 
%         $\frac{1}{2} \left( 1 - \cos \alpha \right)$.}
%\label{fig2}
%\end{figure}
%
%-----------------------------------------------------------------------
% Figure 3:
%-----------------------------------------------------------------------
\clearpage
\newpage

%-----------------------------------------------------------------------
% Figure 4:
%-----------------------------------------------------------------------
\clearpage
\newpage

%-----------------------------------------------------------------------
% Figure 5:
%-----------------------------------------------------------------------
\clearpage
\newpage

%-----------------------------------------------------------------------
% Figure 6:
%-----------------------------------------------------------------------
\clearpage
\newpage

%-----------------------------------------------------------------------
% Figure 7:
%-----------------------------------------------------------------------
\clearpage
\newpage

%-----------------------------------------------------------------------
% Figure 8:
%-----------------------------------------------------------------------
\clearpage
\newpage

%-----------------------------------------------------------------------
% Figure 9:
%-----------------------------------------------------------------------
\clearpage
\newpage

%-----------------------------------------------------------------------
% Figure 10:
%-----------------------------------------------------------------------
\clearpage
\newpage

%-----------------------------------------------------------------------
% Figure 11:
%-----------------------------------------------------------------------
\clearpage
\newpage

%-----------------------------------------------------------------------
% Figure 12:
%-----------------------------------------------------------------------
\clearpage
\newpage

%-----------------------------------------------------------------------
% Figure 13:
%-----------------------------------------------------------------------
\clearpage
\newpage

%-----------------------------------------------------------------------
% Figure 14:
%-----------------------------------------------------------------------
\clearpage
\newpage

%-----------------------------------------------------------------------
% Figure 15:
%-----------------------------------------------------------------------
\clearpage
\newpage

%-----------------------------------------------------------------------
% Figure 16:
%-----------------------------------------------------------------------
\clearpage
\newpage

%-----------------------------------------------------------------------
% Figure 17:
%-----------------------------------------------------------------------
\clearpage
\newpage

%-----------------------------------------------------------------------
% Figure 18:
%-----------------------------------------------------------------------
\clearpage
\newpage

%-----------------------------------------------------------------------
% Table 1:
%-----------------------------------------------------------------------
\clearpage
\newpage

\begin{table}
\vspace*{5.0cm}
\centering
\begin{tabular}{|c|r|r|r|l|l|l|} \hline
nr. & $z$ \hspace*{0.08cm} 
      & $p$ \hspace*{0.2cm} 
      & $T$ \hspace*{0.04cm} 
      & \hspace*{0.5cm} O$_3$ 
      & \hspace*{0.4cm} H$_2$O 
      & \hspace*{0.3cm} $b^{\rm m}_{\rm sca}$ \\ \hline \hline
1 &   0 & 1013.00 &  294 & 0.3041(-1) & 0.1890(+5) & 0.203(-1) \\
2 &   2 &  802.00 &  285 & 0.3712(-1) & 0.9724(+4) & 0.167(-1) \\
3 &   4 &  628.00 &  273 & 0.4830(-1) & 0.3820(+4) & 0.136(-1) \\
4 &   6 &  487.00 &  261 & 0.6420(-1) & 0.1512(+4) & 0.111(-1) \\
5 &   8 &  372.00 &  248 & 0.9126(-1) & 0.6463(+3) & 0.876(-2) \\
6 &  10 &  281.00 &  235 & 0.1306(0) & 0.2475(+3)  & 0.693(-2) \\
7 &  12 &  209.00 &  222 & 0.2216(0) & 0.2952(+2)  & 0.539(-2) \\
8 &  14 &  153.00 &  216 & 0.4409(0) & 0.6526(+1)  & 0.404(-2) \\
9 &  16 &  111.00 &  216 & 0.7053(0) & 0.5727(+1)  & 0.287(-2) \\
10 &  18 &   81.20 &  216 & 0.1295(+1) & 0.6161(+1) & 0.209(-2) \\
11 &  20 &   59.50 &  218 & 0.2171(+1) & 0.7655(+1) & 0.152(-2) \\
12 &  22 &   43.70 &  220 & 0.3162(+1) & 0.1193(+2) & 0.111(-2) \\
13 &  24 &   32.20 &  223 & 0.3852(+1) & 0.1924(+2) & 0.183(-2) \\
14 &  30 &   13.20 &  234 & 0.9131(+1) & 0.4379(+2) & 0.950(-3) \\
15 &  40 &    3.33 &  258 & 0.7431(+1) & 0.2077(+2) & 0.229(-3) \\
16 &  50 & 9.51(-1) & 276 & 0.2728(+1) & 0.1065(+1) & 0.915(-4) \\
17 & 100 & 3.00(-4) & 210 & 0.5191(-1) & 0.3216(+1) & \hspace{0.5cm} - \\ \hline
\end{tabular}
\vspace*{0.5cm}
\caption{The altitude $z$ (in km), the pressure $p$ (in hPa), 
         the temperature $T$ (in K), and the ozone (O$_3$) and
         watervapor (H$_2$O) mixing ratios 
         (in ppm, or parts per million)
         at the 17 levels of the model atmospheres
         \citep{1972McClatchey}.
         Also given is the molecular
         scattering optical thickness $b^{\rm m}_{\rm sca}$
         (at $\lambda=0.55~\mu$m) of each of the 16 {\em layers} of the 
         model atmospheres. The total molecular scattering optical
         thickness of the model atmosphere is 0.0975
         (at $\lambda=0.55~\mu$m).
         Here, $x(-y)$ stands for $x \cdot 10^{-y}$.}
\label{tab1}
\end{table}

%-----------------------------------------------------------------------
\end{document}